\documentclass[iop]{emulateapj}

\usepackage{aas_macros}

\usepackage{color}
\usepackage[dvipsnames,svgnames]{xcolor}

\usepackage{natbib}

\usepackage{amsmath}
\usepackage{xspace}
\usepackage{lineno}

\mathchardef\mhyphen="2D

\newlength{\dhatheight}

\newcommand{\code}[1]{\texttt{#1}\xspace}

\newcommand{\unit}[1]{\ensuremath{\mathrm{\,#1}}\xspace}

\newcommand{\degree}{\ensuremath{{}^{\circ}}\xspace}

\newcommand{\km}{\unit{km}}

\newcommand{\second}{\unit{s}}

\newcommand{\msun}{\unit{M_\odot}}

\newcommand{\lsun}{\unit{L_\odot}}

\providecommand\physrep{\ref@jnl{Phys.~Rep.}}%
\providecommand\apjs{\ref@jnl{ApJS}}%
\providecommand{\jcap}{\ref@jnl{JCAP}}%

\newcommand{\feh}         {\mbox{[Fe/H]}}
\newcommand{\kms}         {\ensuremath{\km~\second^{-1}}\xspace}

\newcommand{\masyr}       {mas~yr$^{-1}$}
\newcommand{\mua}         {\mu_{\alpha}\cos{\delta}}
\newcommand{\mud}         {\mu_{\delta}}

\def\spose#1{\hbox to 0pt{#1\hss}}
\def\lta{\mathrel{\spose{\lower 3pt\hbox{$\mathchar"218$}}
     \raise 2.0pt\hbox{$\mathchar"13C$}}}
\def\gta{\mathrel{\spose{\lower 3pt\hbox{$\mathchar"218$}}
    \raise 2.0pt\hbox{$\mathchar"13E$}}}

\shorttitle{Spectroscopy of Grus~II, Tucana~IV, and Tucana~V}
\shortauthors{Simon et al.}
\slugcomment{Accepted for publication in ApJ}

\begin{document}

\title{Birds of a Feather?  Magellan/IMACS Spectroscopy of the
  Ultra-Faint Satellites Grus~II, Tucana~IV, and Tucana~V{*}}

\altaffiltext{*}{This paper includes data gathered with the 6.5 meter
  Magellan Telescopes located at Las Campanas Observatory, Chile.}


\def\andname{}
\author{
J.~D.~Simon\altaffilmark{1},
T.~S.~Li\altaffilmark{1,2,3,4,$\ddagger$},
D.~Erkal\altaffilmark{5},
A.~B.~Pace\altaffilmark{6,7},
A.~Drlica-Wagner\altaffilmark{3,8,4},
D.~J.~James\altaffilmark{9},
J.~L.~Marshall\altaffilmark{7},
K.~Bechtol\altaffilmark{10,11},
T.~Hansen\altaffilmark{7},
K.~Kuehn\altaffilmark{12,13},
C.~Lidman\altaffilmark{14},
S.~Allam\altaffilmark{3},
J.~Annis\altaffilmark{3},
S.~Avila\altaffilmark{15},
E.~Bertin\altaffilmark{16,17},
D.~Brooks\altaffilmark{18},
D.~L.~Burke\altaffilmark{19,20},
A.~Carnero~Rosell\altaffilmark{21,22},
M.~Carrasco~Kind\altaffilmark{23,24},
J.~Carretero\altaffilmark{25},
L.~N.~da Costa\altaffilmark{22,26},
J.~De~Vicente\altaffilmark{21},
S.~Desai\altaffilmark{27},
P.~Doel\altaffilmark{18},
T.~F.~Eifler\altaffilmark{28,29},
S.~Everett\altaffilmark{30},
P.~Fosalba\altaffilmark{31,32},
J.~Frieman\altaffilmark{3,4},
J.~Garc\'ia-Bellido\altaffilmark{15},
E.~Gaztanaga\altaffilmark{31,32},
D.~W.~Gerdes\altaffilmark{33,34},
D.~Gruen\altaffilmark{35,19,20},
R.~A.~Gruendl\altaffilmark{23,24},
J.~Gschwend\altaffilmark{22,26},
G.~Gutierrez\altaffilmark{3},
D.~L.~Hollowood\altaffilmark{30},
K.~Honscheid\altaffilmark{36,37},
E.~Krause\altaffilmark{28},
N.~Kuropatkin\altaffilmark{3},
N.~MacCrann\altaffilmark{36,37},
M.~A.~G.~Maia\altaffilmark{22,26},
M.~March\altaffilmark{38},
R.~Miquel\altaffilmark{39,25},
A.~Palmese\altaffilmark{3,4},
F.~Paz-Chinch\'{o}n\altaffilmark{23,24},
A.~A.~Plazas\altaffilmark{2},
K.~Reil\altaffilmark{20},
A.~Roodman\altaffilmark{19,20},
E.~Sanchez\altaffilmark{21},
B.~Santiago\altaffilmark{40,22},
V.~Scarpine\altaffilmark{3},
M.~Schubnell\altaffilmark{34},
S.~Serrano\altaffilmark{31,32},
M.~Smith\altaffilmark{41},
E.~Suchyta\altaffilmark{42},
G.~Tarle\altaffilmark{34},
A.~R.~Walker\altaffilmark{43}
\\ \vspace{0.2cm} (DES Collaboration) \\
}
\affil{$^{1}$ Observatories of the Carnegie Institution for Science, 813 Santa Barbara St., Pasadena, CA 91101, USA}
\affil{$^{2}$ Department of Astrophysical Sciences, Princeton University, Peyton Hall, Princeton, NJ 08544, USA}
\affil{$^{3}$ Fermi National Accelerator Laboratory, P. O. Box 500, Batavia, IL 60510, USA}
\affil{$^{4}$ Kavli Institute for Cosmological Physics, University of Chicago, Chicago, IL 60637, USA}
\affil{$^{\ddagger}$  NHFP Einstein Fellow}
\affil{$^{5}$ Department of Physics, University of Surrey, Guildford GU2 7XH, UK}
\affil{$^{6}$ Department of Physics, Carnegie Mellon University, Pittsburgh, Pennsylvania 15312, USA}
\affil{$^{7}$ George P. and Cynthia Woods Mitchell Institute for Fundamental Physics and Astronomy, and Department of Physics and Astronomy, Texas A\&M University, College Station, TX 77843,  USA}
\affil{$^{8}$ Department of Astronomy and Astrophysics, University of Chicago, Chicago, IL 60637, USA}
\affil{$^{9}$ Center for Astrophysics $\vert$ Harvard \& Smithsonian, 60 Garden Street, Cambridge, MA 02138, USA}
\affil{$^{10}$ LSST, 933 North Cherry Avenue, Tucson, AZ 85721, USA}
\affil{$^{11}$ Physics Department, 2320 Chamberlin Hall, University of Wisconsin-Madison, 1150 University Avenue Madison, WI  53706-1390}
\affil{$^{12}$ Australian Astronomical Optics, Macquarie University, North Ryde, NSW 2113, Australia}
\affil{$^{13}$ Lowell Observatory, 1400 Mars Hill Rd, Flagstaff, AZ 86001, USA}
\affil{$^{14}$ The Research School of Astronomy and Astrophysics, Australian National University, ACT 2601, Australia}
\affil{$^{15}$ Instituto de Fisica Teorica UAM/CSIC, Universidad Autonoma de Madrid, 28049 Madrid, Spain}
\affil{$^{16}$ CNRS, UMR 7095, Institut d'Astrophysique de Paris, F-75014, Paris, France}
\affil{$^{17}$ Sorbonne Universit\'es, UPMC Univ Paris 06, UMR 7095, Institut d'Astrophysique de Paris, F-75014, Paris, France}
\affil{$^{18}$ Department of Physics \& Astronomy, University College London, Gower Street, London, WC1E 6BT, UK}
\affil{$^{19}$ Kavli Institute for Particle Astrophysics \& Cosmology, P. O. Box 2450, Stanford University, Stanford, CA 94305, USA}
\affil{$^{20}$ SLAC National Accelerator Laboratory, Menlo Park, CA 94025, USA}
\affil{$^{21}$ Centro de Investigaciones Energ\'eticas, Medioambientales y Tecnol\'ogicas (CIEMAT), Madrid, Spain}
\affil{$^{22}$ Laborat\'orio Interinstitucional de e-Astronomia - LIneA, Rua Gal. Jos\'e Cristino 77, Rio de Janeiro, RJ - 20921-400, Brazil}
\affil{$^{23}$ Department of Astronomy, University of Illinois at Urbana-Champaign, 1002 W. Green Street, Urbana, IL 61801, USA}
\affil{$^{24}$ National Center for Supercomputing Applications, 1205 West Clark St., Urbana, IL 61801, USA}
\affil{$^{25}$ Institut de F\'{\i}sica d'Altes Energies (IFAE), The Barcelona Institute of Science and Technology, Campus UAB, 08193 Bellaterra (Barcelona) Spain}
\affil{$^{26}$ Observat\'orio Nacional, Rua Gal. Jos\'e Cristino 77, Rio de Janeiro, RJ - 20921-400, Brazil}
\affil{$^{27}$ Department of Physics, IIT Hyderabad, Kandi, Telangana 502285, India}
\affil{$^{28}$ Department of Astronomy/Steward Observatory, University of Arizona, 933 North Cherry Avenue, Tucson, AZ 85721-0065, USA}
\affil{$^{29}$ Jet Propulsion Laboratory, California Institute of Technology, 4800 Oak Grove Dr., Pasadena, CA 91109, USA}
\affil{$^{30}$ Santa Cruz Institute for Particle Physics, Santa Cruz, CA 95064, USA}
\affil{$^{31}$ Institut d'Estudis Espacials de Catalunya (IEEC), 08034 Barcelona, Spain}
\affil{$^{32}$ Institute of Space Sciences (ICE, CSIC),  Campus UAB, Carrer de Can Magrans, s/n,  08193 Barcelona, Spain}
\affil{$^{33}$ Department of Astronomy, University of Michigan, Ann Arbor, MI 48109, USA}
\affil{$^{34}$ Department of Physics, University of Michigan, Ann Arbor, MI 48109, USA}
\affil{$^{35}$ Department of Physics, Stanford University, 382 Via Pueblo Mall, Stanford, CA 94305, USA}
\affil{$^{36}$ Center for Cosmology and Astro-Particle Physics, The Ohio State University, Columbus, OH 43210, USA}
\affil{$^{37}$ Department of Physics, The Ohio State University, Columbus, OH 43210, USA}
\affil{$^{38}$ Department of Physics and Astronomy, University of Pennsylvania, Philadelphia, PA 19104, USA}
\affil{$^{39}$ Instituci\'o Catalana de Recerca i Estudis Avan\c{c}ats, E-08010 Barcelona, Spain}
\affil{$^{40}$ Instituto de F\'\i sica, UFRGS, Caixa Postal 15051, Porto Alegre, RS - 91501-970, Brazil}
\affil{$^{41}$ School of Physics and Astronomy, University of Southampton,  Southampton, SO17 1BJ, UK}
\affil{$^{42}$ Computer Science and Mathematics Division, Oak Ridge National Laboratory, Oak Ridge, TN 37831}
\affil{$^{43}$ Cerro Tololo Inter-American Observatory, National Optical Astronomy Observatory, Casilla 603, La Serena, Chile}

\begin{abstract}
We present Magellan/IMACS spectroscopy of three recently discovered
ultra-faint Milky Way satellites, Grus~II, Tucana~IV, and Tucana~V.
We measure systemic velocities of $v_{hel} = -110.0 \pm 0.5$~\kms,
$v_{hel} = 15.9^{+1.8}_{-1.7}$~\kms, and $v_{hel} =
-36.2^{+2.5}_{-2.2}$~\kms for the three objects, respectively.  Their
large relative velocities demonstrate that the satellites are
unrelated despite their close physical proximity.  We determine a
velocity dispersion for Tuc~IV of $\sigma = 4.3^{+1.7}_{-1.0}$~\kms,
but we cannot resolve the velocity dispersions of the other two
systems.  For Gru~II we place an upper limit (90\%\ confidence) on the
dispersion of $\sigma < 1.9$~\kms, and for Tuc~V we do not obtain any
useful limits.  All three satellites have metallicities below
$\mbox{[Fe/H]} = -2.1$, but none has a detectable metallicity spread.
We determine proper motions for each satellite based on \emph{Gaia}
astrometry and compute their orbits around the Milky Way.  Gru~II is
on a tightly bound orbit with a pericenter of $25^{+6}_{-7}$~kpc and
orbital eccentricity of $0.45^{+0.08}_{-0.05}$.  Tuc~V likely has an
apocenter beyond 100~kpc, and could be approaching the Milky Way for
the first time.  The current orbit of Tuc~IV is similar to that of
Gru~II, with a pericenter of $25^{+11}_{-8}$~kpc and an eccentricity
of $0.36^{+0.13}_{-0.06}$.  However, a backward integration of the
position of Tuc~IV demonstrates that it collided with the Large
Magellanic Cloud at an impact parameter of $4$~kpc $\sim120$~Myr
ago, deflecting its trajectory and possibly altering its internal
kinematics.  Based on their sizes, masses, and metallicities, we
classify Gru~II and Tuc~IV as likely dwarf galaxies, but the nature of
Tuc~V remains uncertain.
\end{abstract}



\section{INTRODUCTION}
\label{intro}

Over the past four years, analyses of the first deep, wide-field
digital surveys of the southern sky have significantly expanded the
population of Milky Way satellites and pushed the search for faint
dwarf galaxies to lower surface brightnesses than was previously
possible
\citep{bechtol15,koposov15,koposov18,drlica15,dw16,laevens15,laevens15b,
  kim15peg3,kj15,torrealba16,torrealba16b,torrealba18}.  Motivated by
the apparent concentration of satellites near the Large and Small
Magellanic Clouds, one particularly active area of recent study is the
association of dwarf galaxies with the Magellanic system
\citep[e.g.,][]{sales11,deason15,jethwa16,sales17,kallivayalil18,pardy19,eb19}.
Dwarfs may also be associated with each other in even smaller groups
\citep[e.g.,][]{lh08,klimentowski10}.

Several of the recently discovered Milky Way satellites are in very
close physical proximity to each other, sparking the suggestion that
they could represent bound (or formerly bound) associations of dwarfs.
Carina~II and Carina~III are just 8~kpc apart at present
\citep{torrealba18}, but their radial velocities \citep{li18} and
orbits \citep{simon18} are so different that this configuration
appears to be simply a coincidence.  Similarly, \citet{drlica15} noted
that among the satellites discovered by the Dark Energy Survey
\citep[DES;][]{diehl14,decam}, Tucana~II, Tucana~IV (Tuc~IV), and
Tucana~V (Tuc~V) are all within 7~kpc of a common centroid.  Grus~II
(Gru~II) is located less than 18~kpc from this position as well.  The
radial velocity of Tucana~II has been measured \citep{walker16}, but
no spectroscopy is available for the other three systems, so the
viability of their association with each other or with the Magellanic
Clouds has not been tested.  Finding dwarf galaxies that formed with
only other dwarfs as their neighbors will enable novel tests of the
effects of environment on early galaxy formation and evolution
\citep[e.g.,][]{dl08,wetzel15,rw19,fillingham19}.

This paper continues a series of publications aimed at characterizing
recently discovered Milky Way satellites
\citep{simon15,simon17,li17,li18}.  Here we present the first stellar
spectroscopy of Gru~II, Tuc~IV, and Tuc~V and assess the nature of
each object.  In Section~\ref{observations} we describe our
observations and data reduction.  We measure radial velocities and
metallicities of the observed stars and identify members of the three
satellites in Section~\ref{measurements}.  In Section~\ref{results} we
calculate their dynamical masses, metallicity distributions, proper
motions, and orbits around the Milky Way.  We discuss the nature and
origin of each satellite in Section~\ref{sec:discussion} and summarize
our findings in Section~\ref{conclusions}.

\section{OBSERVATIONS AND DATA REDUCTION}
\label{observations}

\subsection{Spectrograph Setup and Summary of Observations}

We observed Gru~II, Tuc~IV, and Tuc~V with the IMACS spectrograph
\citep{dressler06} on the Magellan/Baade telescope during a number of
observing runs spanning from 2015 August to 2018 September.  As in
previous studies \citep{simon17,li17,li18} we used the f/4 camera and
the 1200~$\ell$/mm grating blazed at 9000~\AA\ to produce $R \approx
11,000$ spectra from $\sim7500-8900$~\AA.  We observed 7 slit masks
targeting Gru~II, 9 slit masks targeting Tuc~IV, and 1 slit mask
targeting Tuc~V.  Each mask was designed with $0\farcs7 \times
5\arcsec$ slitlets.  Typically, our observing sequence consisted of
several science exposures totaling $1-2$~hours followed by comparison
lamp and flat field frames.  The comparison lamps used were Ne, Ar,
and He through 2015 October, Ne, Ar, and Kr from 2015 November until
2016 August, and Ne, Ar, Kr, and He beginning in 2017.  Total
integration times for each slit mask ranged from 0.5~hr to
$\sim10$~hr.  A summary of the observing dates and mask parameters is
provided in Table~\ref{obstable}.

\begin{deluxetable*}{lcccrccc}
\tablecaption{Summary of IMACS Spectroscopic Observations}
\tablewidth{0pt}
\tablehead{
\colhead{Mask} &
\colhead{$\alpha$ (J2000)} &
\colhead{$\delta$ (J2000)} &
\colhead{Slit PA} &
\colhead{$t_{\rm exp}$} &
\colhead{MJD of} &
\colhead{\# of slits} &
\colhead{\% useful} \\
\colhead{name}&
\colhead{(h$\,$ $\,$m$\,$ $\,$s)} &
\colhead{($^\circ\,$ $\,'\,$ $\,''$)} &
\colhead{(deg)} &
\colhead{(sec)} &
\colhead{observation\tablenotemark{a}} &
\colhead{} &
\colhead{spectra}
}
\startdata
Gru II Mask 1     & 22 03 46.80 & $-46$ 28 50.0  & 240.0 & 20400    & 57282.59 & 75 & \phn41\%\\
                  &             &                &       & \phn7200 & 57634.36 &    & \phn20\%\\
Gru II Mask 2     & 22 04 04.00 & $-46$ 30 30.0  & 316.0 & 13800    & 57311.28 & 71 & \phn37\%\\
Gru II Mask 3     & 22 04 00.00 & $-46$ 19 00.0  & 300.0 & \phn7200 & 57284.08 & 59 & \phn42\%\\
                  &             &                &       & \phn5400 & 57630.53 &    & \phn42\%\\
Gru II Mask 4     & 22 03 57.00 & $-46$ 27 47.0  & 260.0 & 34800    & 57980.89 & 67 & \phn45\%\\
Gru II Mask 5     & 22 04 45.00 & $-46$ 30 20.0  & 195.0 & \phn4800 & 57981.04 & 61 & \phn25\%\\
Gru II Mask 6     & 22 04 27.00 & $-46$ 26 20.0  & 282.0 & \phn9000 & 58339.54 & 48 & \phn77\%\\
Gru II Mask 7     & 22 03 37.00 & $-46$ 27 25.0  & 134.0 & \phn7920 & 58363.60 & 50 & \phn40\%\\
Tuc IV Mask 1     & 00 03 02.20 & $-60$ 49 10.0  & 226.0 & \phn9600 & 57247.35 & 58 & \phn45\%\\
                  &             &                &       & \phn9600 & 57282.32 &    & \phn47\%\\
                  &             &                &       & \phn9600 & 57346.11 &    & \phn52\%\\
Tuc IV Mask 2     & 00 02 44.00 & $-60$ 49 27.0  & 166.2 & 15480    & 57633.91 & 66 & \phn26\%\\
Tuc IV Mask 3     & 00 01 51.00 & $-60$ 55 30.0  & 154.0 & \phn4800 & 57981.23 & 44 & \phn34\%\\
Tuc IV Mask 4     & 00 04 53.00 & $-60$ 46 00.0  & 192.0 & \phn3600 & 57982.14 & 52 & \phn38\%\\
Tuc IV Mask 5     & 00 02 55.00 & $-60$ 47 30.0  & 202.0 & \phn9600 & 58339.18 & 45 & \phn60\%\\
Tuc IV Mask 6     & 00 02 14.50 & $-61$ 04 07.0  & 269.0 & 16680    & 58340.26 & 29 & \phn72\%\\
Tuc IV Mask 7     & 00 02 19.30 & $-60$ 41 28.0  & 265.0 & \phn9600 & 58339.32 & 38 & \phn61\%\\
Tuc IV Mask 8     & 00 02 59.00 & $-61$ 09 00.0  & 145.0 & \phn1800 & 58340.13 &  5 & \phn40\%\\
Tuc IV Mask 9     & 00 02 13.00 & $-60$ 46 15.0  & 134.0 & \phn9600 & 58363.75 &  8 & \phn63\%\\
Tuc V Mask 1      & 23 37 40.00 & $-63$ 16 34.0  & 246.0 & \phn3840 & 57283.32 & 65 & \phn17\%\\
                  &             &                &       & \phn9600 & 57312.23 &    & \phn23\%\\
                  &             &                &       &  33840   & 57568.70 &    & \phn43\%
\enddata
\tablenotetext{a}{For observations made over multiple nights, the date listed here is 
the weighted mean observation date, which may occur during daylight hours.  }
\label{obstable}
\end{deluxetable*}

\subsection{Target Selection}
\label{sec:targets}

Spectroscopic targets were selected from the DES Y2Q1 photometric
catalog.  The target selection criteria were identical to those used
by \citet{simon17} for Tucana~III.  Candidate red giant branch (RGB)
stars were selected in a window defined by a 12~Gyr, $\feh = -2.20$
PARSEC isochrone \citep{bressan12} on the red side and the
\citet{an08} M92 fiducial sequence (transformed from SDSS to the DES
photometric system) on the blue side.\footnote{Note that the PARSEC
  isochrones relied on outdated throughput curves for the DES filters
  at the time of the target selection for this paper.  This problem
  was corrected in 2017 July (see appendix of \citealt{li18}), so
  isochrones downloaded now will not match those used for our target
  selection.  Nevertheless, because our color selection window was
  defined to be wider than the giant branch of known ultra-faint
  dwarfs, use of the old isochrones should not bias our spectroscopic
  target selection.}  Stars outside this window but within 0.03~mag of
either edge were targeted at reduced priority.  Candidate horizontal
branch stars in these systems are located in a relatively clean part
of the color-magnitude diagram (CMD) and were selected according to
the photometric membership probabilities from \citet{drlica15}.  We
also targeted main sequence turnoff stars based on their membership
probabilities.

In 2016 July we obtained some shallow spectroscopy of wide fields
surrounding Gru~II and Tuc~IV with the AAOmega spectrograph
\citep{sharp06} on the Anglo-Australian Telescope (AAT).  These
observations took place in poor conditions and served primarily to
measure the velocities of bright stars to see if any could be
associated with Gru~II or Tuc~IV.  We identified one new member of
Gru~II in this data set (which we subsequently confirmed with IMACS)
and eliminated many stars in the photometric selection region from
further consideration.  Targeting for the 2017 and 2018 IMACS
observations (Gru~II Masks 4-7 and Tuc~IV Masks 4-9) employed this
information.  Although the AAT velocity measurements are not used in
our analysis, we provide the data in Table~\ref{tab:aat} for the
benefit of any other researchers interested in these fields.


\tabletypesize{\scriptsize}
\begin{deluxetable*}{c c c c c c c r}[t!]
\tablecaption{AAT Velocity Measurements for Gru~II and Tuc~IV.\label{tab:aat}}

\tablehead{ID & MJD & RA & DEC & $g$\tablenotemark{a} & $r$\tablenotemark{a} & S/N & \multicolumn{1}{c}{$v$} \\ 
 &  & (deg) & (deg) & (mag) & (mag) &  & \multicolumn{1}{c}{(\kms)} }
\startdata
DES\,J215828.76$-$463329.6  &  57596.1  &  329.61985  & $ -46.55823 $ &  17.73  &  17.03  &    6.8  &    \phs$4.8 \pm 3.5$  \\
DES\,J215831.30$-$462702.6  &  57596.1  &  329.63040  & $ -46.45072 $ &  17.22  &  16.43  &   17.8  &      $-11.4 \pm 2.2$  \\
DES\,J215851.85$-$464030.5  &  57596.1  &  329.71605  & $ -46.67514 $ &  17.90  &  17.23  &    7.3  &      $-62.9 \pm 3.3$  \\
DES\,J215857.30$-$460420.0  &  57596.1  &  329.73874  & $ -46.07222 $ &  17.06  &  16.13  &   29.0  &      $-12.5 \pm 0.8$  \\
DES\,J215903.05$-$465407.6  &  57596.1  &  329.76272  & $ -46.90211 $ &  17.06  &  16.23  &    8.6  &      $-37.0 \pm 3.0$  \\
DES\,J215909.17$-$463536.3  &  57596.1  &  329.78819  & $ -46.59340 $ &  17.78  &  17.11  &    9.0  &       $-7.6 \pm 3.1$  \\
DES\,J215915.51$-$461956.4  &  57596.1  &  329.81464  & $ -46.33234 $ &  17.10  &  16.28  &   26.5  &      $-17.4 \pm 2.6$  \\
DES\,J215917.67$-$465605.9  &  57596.1  &  329.82361  & $ -46.93498 $ &  17.31  &  16.56  &    9.8  &   \phs$41.2 \pm 3.9$  \\
DES\,J215923.64$-$460638.6  &  57596.1  &  329.84849  & $ -46.11072 $ &  17.19  &  16.38  &   26.4  &      $-16.1 \pm 1.5$  \\
DES\,J215924.02$-$461309.8  &  57596.1  &  329.85008  & $ -46.21940 $ &  17.57  &  16.77  &   11.4  &   \phs$54.2 \pm 1.9$  \\
\enddata
\tablenotetext{a}{Quoted magnitudes represent the weighted-average dereddened PSF magnitude derived from the DES images using SourceExtractor \citep{drlica15}.}
\tablecomments{This table is available in its entirety in the electronic edition of the journal.  A portion is reproduced here to provide guidance on form and content.}
\end{deluxetable*}

\begin{figure*}[th!]
\epsscale{1.2}
\plotone{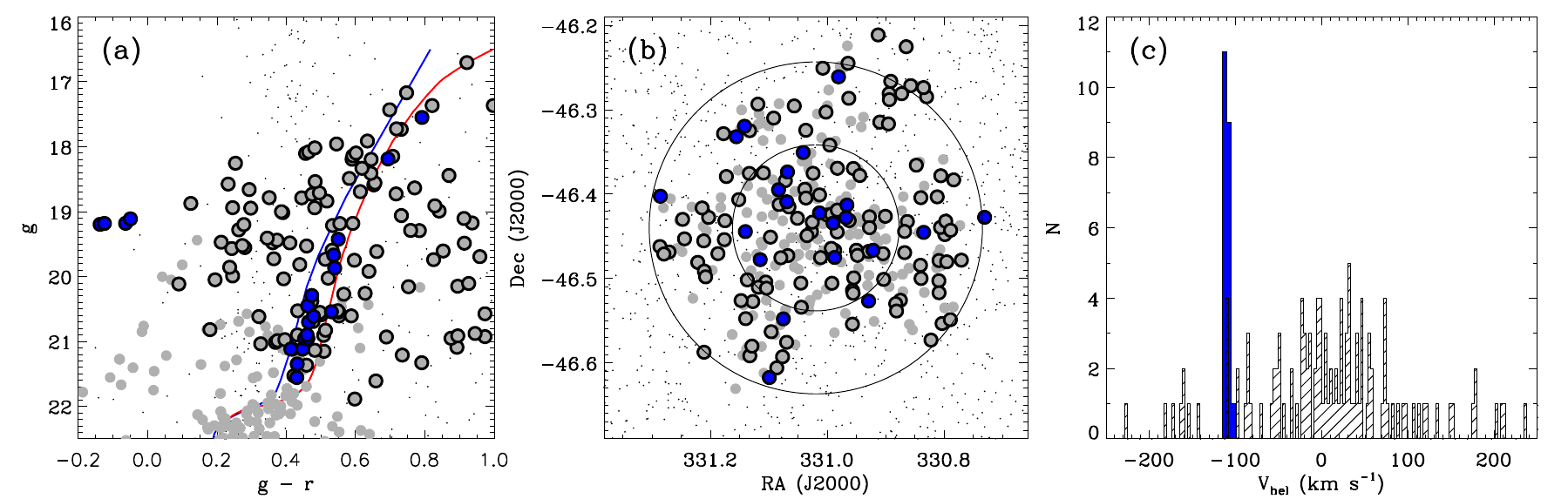}
\caption{\emph{(a)} DES color-magnitude diagram of Grus~II.  Stars
  within 12\arcmin\ of the center of Gru~II are plotted as small
  black dots, and stars selected for spectroscopy (as described in
  \S\ref{sec:targets}) are plotted as filled gray circles.  Points
  surrounded by black outlines represent the stars for which we
  obtained successful velocity measurements, and those we identify as
  Tuc~III members are filled in with blue.  The M92 sequence and PARSEC
  isochrone used to define the RGB of Gru~II are displayed as blue
  and red curves, respectively.  \emph{(b)} Spatial distribution of
  the observed stars.  Symbols are as in panel \emph{(a)}.  Because
  the ellipticity of Gru~II is unconstrained by presently available
  data \citep{drlica15}, we represent the half-light radius of Gru~II
  with a black circle.  The larger circle indicates twice the
  half-light radius.  \emph{(c)} Radial velocity distribution of
  observed stars.  The clear narrow peak of stars at $v \sim
  -100$~\kms highlighted in blue is the signature of Gru~II.  The
  hatched histogram indicates stars that are not members of Gru~II.
{\vskip0.1in}
}
\label{cmd_gru2}
\end{figure*}

\begin{figure*}[t!]
\epsscale{1.2}
\plotone{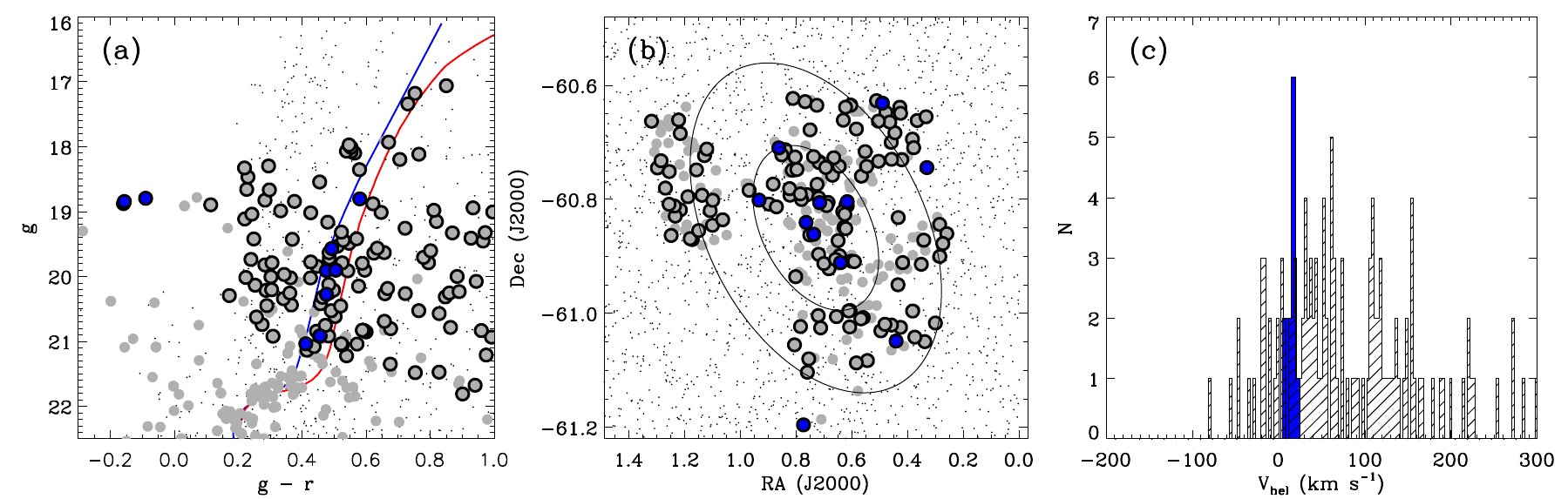}
\caption{\emph{(a)} DES color-magnitude diagram of Tucana~IV.  Stars
  within 17\arcmin\ of the center of Tuc~IV are plotted as small black
  dots, and stars selected for spectroscopy (as described in
  \S\ref{sec:targets}) are plotted as filled gray circles.  Points
  surrounded by black outlines represent the stars for which we
  obtained successful velocity measurements, and those we identify as
  Tuc~IV members are filled in with blue.  The M92 sequence and PARSEC
  isochrone used to define the RGB are displayed as blue and red
  curves, respectively.  \emph{(b)} Spatial distribution of the
  observed stars.  Symbols are as in panel \emph{(a)}.  The black
  ellipses represent one and two times the half-light radius of
  Tuc~IV.  \emph{(c)} Radial velocity distribution of observed stars.
  Tuc~IV members are highlighted in blue, and the hatched histogram
  indicates stars that are not members of Tuc~IV.}
\label{cmd_tuc4}
\end{figure*}

\begin{figure*}[th!]
\epsscale{1.2}
\plotone{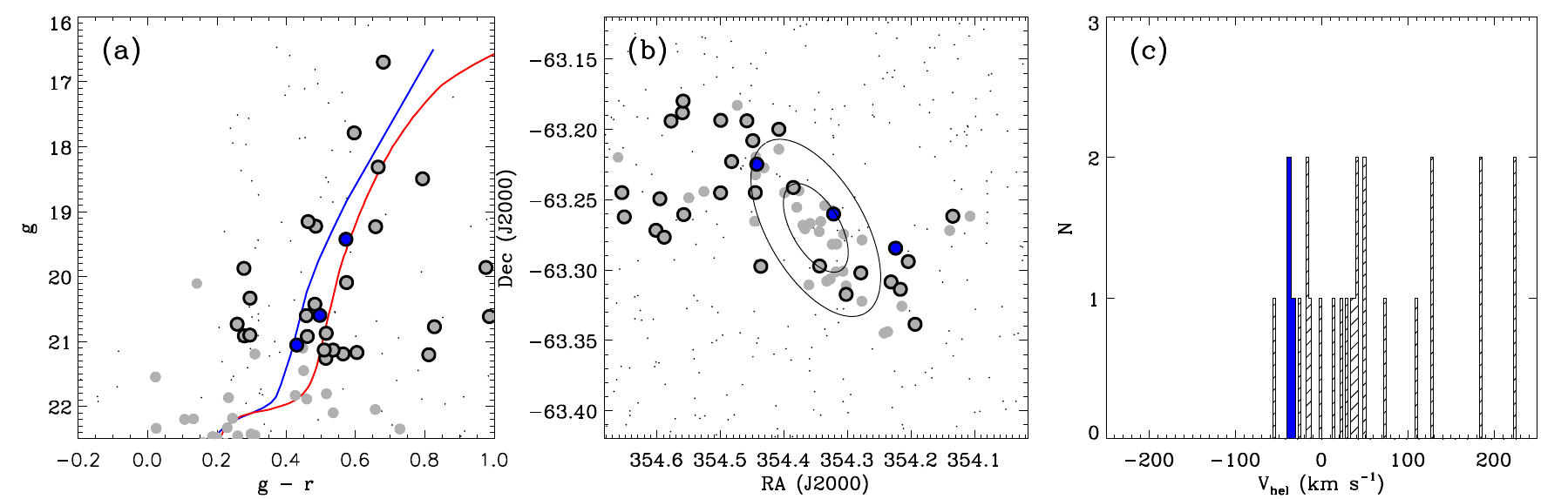}
\caption{\emph{(a)} DES color-magnitude diagram of Tucana~V.  Stars
  within 7\arcmin\ of the center of Tuc~V are plotted as small black
  dots, and stars selected for spectroscopy (as described in
  \S\ref{sec:targets}) are plotted as filled gray circles.  Points
  surrounded by black outlines represent the stars for which we
  obtained successful velocity measurements, and those we identify as
  Tuc~V members are filled in with blue.  The M92 sequence and PARSEC
  isochrone used to define the RGB are displayed as blue and red
  curves, respectively.  \emph{(b)} Spatial distribution of the
  observed stars.  Symbols are as in panel \emph{(a)}.  The black
  ellipses represent one and two times the half-light radius of Tuc~V.
  \emph{(c)} Radial velocity distribution of observed stars.  Tuc~V
  members are highlighted in blue, and the hatched histogram indicates
  stars that are not members of Tuc~V.}
\label{cmd_tuc5}
\end{figure*}

For the 2018 IMACS observations (Gru~II Masks 6-7 and Tuc~IV Masks
5-9) we also included astrometry from the second Gaia data release
(DR2; \citealt{gaiadr2brown}) as an additional (loose) selection
criterion.  Stars with parallaxes differing from zero at less than
$2\sigma$ significance and with proper motions within $3\sigma$ of the
mean proper motion established by the member stars we had already
identified to that point were considered candidate members.  The
color-magnitude and spatial distributions of target stars are
displayed in the left and middle panels, respectively, of
Figures~\ref{cmd_gru2}, \ref{cmd_tuc4}, and \ref{cmd_tuc5}.

\subsection{Data Reduction}
\label{sec:reductions}

We reduced the IMACS observations using the procedures described by
\citet{simon17} and \citet{li17}.  Briefly, we used the Cosmos
reduction pipeline \citep{dressler11,oemler17} to create a map of the
slits across the detector mosaic and generate a preliminary wavelength
solution.  We then employed the IMACS reduction pipeline developed by
\citet{simon17}, which is based on the DEEP2 data reduction pipeline
\citep{cooper12,newman13}, to calibrate and extract one-dimensional
spectra from the raw two-dimensional data.  We note that the IMACS
detector arrays were swapped between the f/2 and f/4 cameras in 2017
December.  The CCD mosaic used in f/4 beginning in 2018 has slightly
higher efficiency at the wavelength of our observations, at the cost
of increased cosmetic artifacts and fringing.  The artifacts can be
removed by slightly changing the grating angle between sets of
observations, and the fringing can be removed via flatfield frames.

\section{VELOCITY AND METALLICITY MEASUREMENTS}
\label{measurements}

\subsection{Radial Velocity Measurements}
\label{sec:velocities}

We measure radial velocities from the reduced IMACS spectra following
the procedures described by \citet{simon17} and \citet{li17}.  We
compute $\chi^{2}$ for each spectrum as a function of velocity
relative to a template spectrum of a bright metal-poor star.  The
template spectra were obtained with IMACS using the same spectrograph
setup but driving the star across the spectrograph slit during the
exposure in order to create a uniform slit illumination profile.  For
RGB candidates we use HD~122563 as the template star, and we assume
that its velocity is $v_{hel} = -26.51$~\kms\ \citep{chubak12}.  For
horizontal branch candidates we use HD~161817 as the template star;
while the spectrum of this hot star is dominated by the hydrogen
Paschen series lines, the Ca triplet lines are visible as well, and we
use them to tie the velocity of the template spectrum to that of
HD~122563.  We measure a velocity of $v_{hel} = -363.27$~\kms\ for
HD~161817, identical within the uncertainties to the value of $v_{hel}
= -363.2 \pm 0.4$~\kms\ determined by \citet{gontcharov06}.  We remove
velocity offsets resulting from imperfect centering of each star in
its slitlet using template fits of the A-band region of each spectrum
to a spectrum of the hot, rapidly rotating star HR~4781.

As in previous papers, because of the inferior wavelength solutions
obtained without the Kr comparison lamp the systematic velocity
uncertainty is 1.2~\kms\ for observations obtained through 2015
October and 1.0~\kms\ beginning in 2015 November.  The statistical
uncertainty on each velocity measurement is determined by adding
normally distributed noise to the observed spectrum and remeasuring
the velocity 500 times \citep{sg07,simon17}.  The total velocity
uncertainties consist of the quadrature sum of the statistical and
systematic uncertainties.

\subsection{Metallicity Measurements}
\label{metallicity_measurements}

We measure metallicities from the equivalent widths of the Ca triplet
(CaT) lines using the methods discussed in \citet{simon15,simon17} and
\citet{li17}.  We fit the CaT lines using a Gaussian plus Lorentzian
profile and integrate the fitted profile to determine the equivalent
width (EW) of each of the three lines.  We assume a systematic
uncertainty on EW measurements of 0.32~\AA\ \citep{simon17}, which we
add in quadrature with the measurement uncertainty to obtain a total
uncertainty.  We convert the summed EWs to metallicity with the
calibration of \citet{carrera13}.  Although Gru~II and Tuc~IV do
contain a handful of HB stars, unlike Reticulum~II and Tucana~III, for
consistency with past work (and with Tuc~V, which lacks any HB stars)
we continue to employ the metallicity calibration based on absolute V
magnitude.  If we used the $V - V_{HB}$ calibration from
\citet{carrera13} instead, we would derive lower metallicities by
$\sim0.1$~dex on average.

\subsection{Spectroscopic Membership Determination}
\label{sec:membership}

\begin{figure*}[th!]
\epsscale{1.2}
\plotone{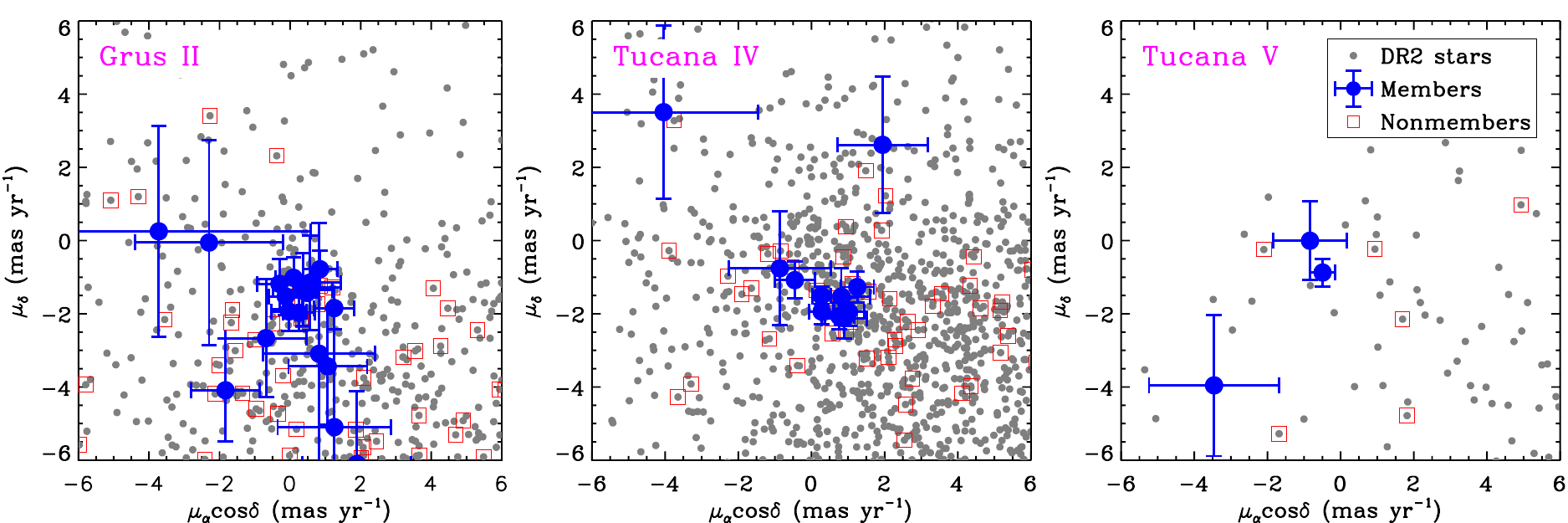}
\caption{Proper motion diagrams for Gru~II (left), Tuc~IV (middle),
  and Tuc~V (right).  The identified members of each satellite are
  plotted as filled blue circles, while spectroscopic non-members are
  shown as open red squares.  The small gray points represent other
  stars in the \emph{Gaia} DR2 catalog within $3r_{1/2}$ of each
  system for which we did not obtain spectra.  The tight clustering of
  the member stars in proper motion is evident for Gru~II and Tuc~IV.
  Two faint members of Tuc~IV and one of Tuc~V are near the magnitude
  limit of the \emph{Gaia} catalog and appear as outliers here, but
  given the large astrometric uncertainties their proper motions are
  within $\sim2\sigma$ of the mean value.}
\label{pmfig}
\end{figure*}

\begin{figure*}[th!]
\epsscale{0.8}
\plotone{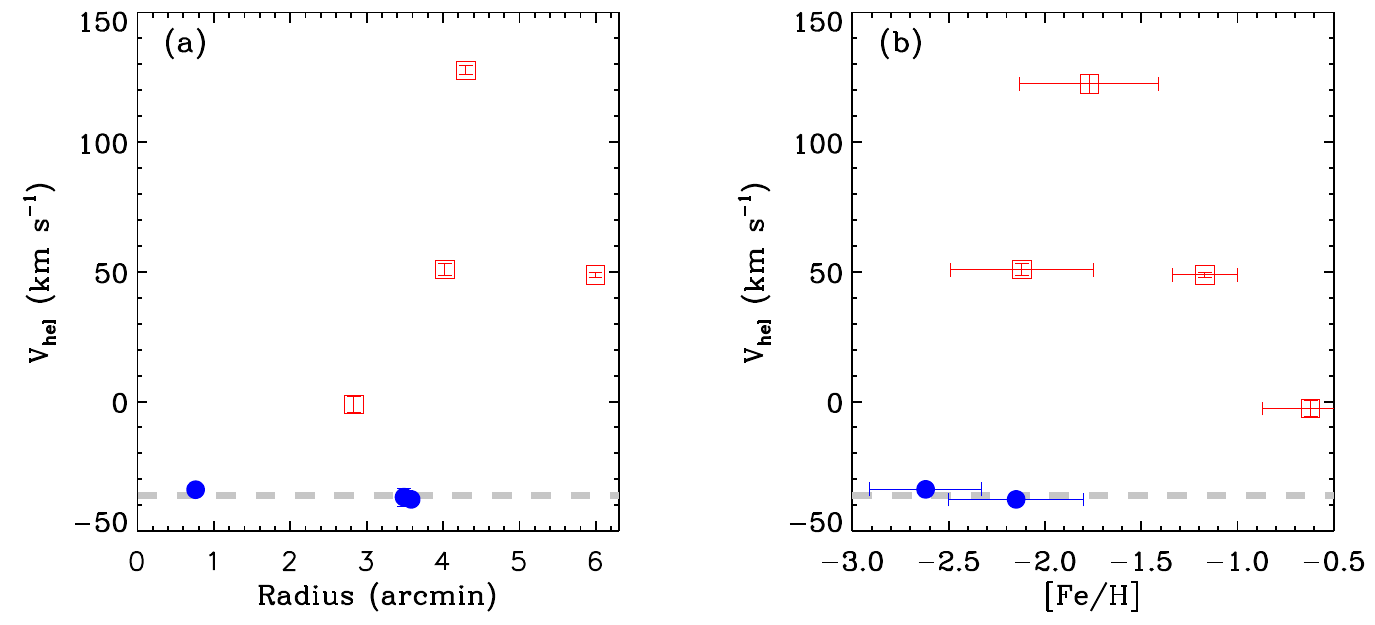}
\caption{(left) Velocity as a function of distance away from the
  center of Tuc~V for stars in the RGB selection region shown in
  Fig.~\ref{cmd_tuc5}a.  The identified Tuc~V members are plotted as
  filled blue circles, while the spectroscopic non-members are shown
  as open red squares.  The systemic velocity of Tuc~V is displayed as
  a dashed gray line. (right) Velocity as a function of metallicity
  for stars in the RGB selection region (between the two isochrones
  and brighter than $g \approx 21.1$) shown in Fig.~\ref{cmd_tuc5}a.
  Symbols are as in the left panel.  The spectrum of the faintest
  Tuc~V member does not have a high enough S/N ratio for an accurate
  measurement of the CaT equivalent width, so only the two brighter
  members are plotted in this panel.  Note that because CaT
  metallicities depend on the assumed distance, the metallicities for
  the non-Tuc~V stars are likely underestimated (presuming that these
  stars are in the foreground).  The Tuc~V members share a common
  velocity and are closer to the center and more metal-poor than the
  non-members.}
\label{tuc5plots}
\end{figure*}

For each satellite, we determine an approximate value for the systemic
velocity by considering the most likely member candidates (stars lying
along the fiducial sequences determined from previously studied DES
dwarfs and closest to the center of the object) from the initial
target selection.  We then select stars within the various photometric
selection regions and with velocities within 15~\kms\ of the systemic
velocity as member candidates.  The candidates are inspected
individually, considering their precise position in the CMD,
metallicity (if available), and other spectral diagnostics such as the
\ion{Mg}{1}~$\lambda8807$~\AA\ line \citep{bs12}.  Finally, we
cross-match the resulting set of stars with the \emph{Gaia} DR2
catalog.  As with the other ultra-faint satellites \citep{simon18},
the member stars of Gru~II and Tuc~IV are tightly clustered in proper
motion space \citep{mh18}, and any outliers are immediately obvious
(see Figure~\ref{pmfig}).
One candidate Tuc~IV member is rejected on the basis of \emph{Gaia}
astrometry.  Because Tuc~V only contains a single member candidate
brighter than $g=20$, the \emph{Gaia} proper motions do not provide as
much discriminatory power, but we verify that the fainter candidates
have proper motions consistent with the brighter one despite their
large uncertainties.  In particular, three of the five
highest-priority Tuc~V target stars are clustered at a velocity of
$\sim-36$~\kms, confirming the detection of the satellite despite the
small sample (see Figure~\ref{tuc5plots}).

After combining the available photometric data, radial velocities,
metallicities, spectral diagnostics, and astrometry, most of the
members of each system are unambiguous.  The exceptions are one star
each in Gru~II (DES~J220400.12$-$462529.0) and Tuc~IV
(DES~J000311.46$-$604451.5), which resemble members in every way
except that their color places them $\sim0.1$~mag redder than the RGB
at their magnitude.  Conceivably these stars could be members that are
carbon-enhanced, similar to one star in Hydrus~I identified by
\citet{koposov18} and another star in Canes~Venatici~I studied by
\citet{zucker06b} and \citet{yoon19}.  Additional spectroscopy at
shorter wavelengths where carbon features are prominent would be
needed to test this possibility.  However, given the currently
available data for these stars we consider them non-members.  In both
cases, because their properties match the mean properties of the
satellites so closely, all of our conclusions would be unaffected if
they were included as members.  A second Tuc~IV candidate
(DES~J000244.67$-$604819.1) that is on the RGB in the CMD and located
near the center of the system is a $\sim2\sigma$ outlier in both
velocity and proper motion, as well as having a relatively high CaT
metallicity of $\feh = -1.43$, so we judge it a non-member.  If it
were a member of Tuc~IV it would increase the inferred velocity
dispersion by less than $1\sigma$.  Our spectroscopic data sets
include 21 members of Gru~II, 11 members of Tuc~IV, and 3 members of
Tuc~V.

All Magellan/IMACS velocity and metallicity measurements, as well as
membership determinations, are listed in Table~\ref{tab:imacs_spec_table}.


\tabletypesize{\scriptsize}
\begin{deluxetable*}{c c c c c c c r c c c}
\tablecaption{IMACS Velocity and Metallicity Measurements for Gru~II, Tuc~IV, and Tuc~V.\label{tab:imacs_spec_table}}

\tablehead{ID & RA & DEC & $g$\tablenotemark{a} & $r$\tablenotemark{a} & MJD & S/N & \multicolumn{1}{c}{$v$} & ${\rm EW}$ & ${\rm [Fe/H]}$ & Mem\tablenotemark{b} \\ 
 & (deg) & (deg) & (mag) & (mag) &  &  & \multicolumn{1}{c}{(\kms)} & (\AA) &  &  }
\startdata
DES\,J220255.41$-$462538.9  &  330.73088  & $ -46.42747 $ &  19.18  &  19.24  &  58363.6  &    5.9  &     $-104.4 \pm 3.3$  &         ...       &         ...        &  1 \\
DES\,J220304.94$-$462841.7  &  330.77059  & $ -46.47824 $ &  20.59  &  20.09  &  57282.6  &   11.8  &     $-142.7 \pm 4.1$  &  $5.98 \pm 0.61$  &         ...        &  0 \\
DES\,J220308.10$-$462259.2  &  330.78377  & $ -46.38312 $ &  20.15  &  19.25  &  58363.6  &    8.6  &  \phs$107.3 \pm 2.7$  &         ...       &         ...        &  0 \\
DES\,J220309.37$-$462634.2  &  330.78903  & $ -46.44283 $ &  18.44  &  17.57  &  58363.6  &   34.5  &   \phs$23.4 \pm 1.2$  &  $5.65 \pm 0.38$  &         ...        &  0 \\
DES\,J220309.47$-$463255.6  &  330.78944  & $ -46.54878 $ &  18.69  &  18.08  &  57282.6  &   45.7  &   \phs$40.4 \pm 1.2$  &  $5.75 \pm 0.36$  &         ...        &  0 \\
                            &  330.78944  & $ -46.54878 $ &  18.69  &  18.08  &  57634.4  &   18.4  &   \phs$41.8 \pm 1.5$  &  $5.76 \pm 0.49$  &         ...        &  0 \\
DES\,J220310.98$-$462848.8  &  330.79573  & $ -46.48023 $ &  18.78  &  18.43  &  58363.6  &   16.6  &   \phs$42.2 \pm 1.4$  &  $3.95 \pm 0.45$  &         ...        &  0 \\
DES\,J220311.39$-$462555.3  &  330.79746  & $ -46.43203 $ &  19.48  &  19.12  &  58363.6  &    8.3  &      $-18.3 \pm 1.7$  &  $2.56 \pm 0.55$  &         ...        &  0 \\
DES\,J220311.75$-$462652.7  &  330.79897  & $ -46.44796 $ &  21.15  &  20.64  &  57980.9  &    5.6  &   \phs$69.2 \pm 3.2$  &  $4.01 \pm 1.81$  &         ...        &  0 \\
DES\,J220312.90$-$463312.1  &  330.80374  & $ -46.55336 $ &  19.61  &  18.95  &  57282.6  &   26.1  &      $-13.9 \pm 1.4$  &  $5.75 \pm 0.41$  &         ...        &  0 \\
                            &  330.80374  & $ -46.55336 $ &  19.61  &  18.95  &  57634.4  &    8.4  &      $-16.0 \pm 2.9$  &         ...       &         ...        &  0 \\
\enddata
\tablenotetext{a}{Quoted magnitudes represent the weighted-average dereddened PSF magnitude derived from the DES images using SourceExtractor \citep{drlica15}.}
\tablenotetext{b}{A value of 1 indicates that the star is a member of the relevant satellite, while 0 indicates a non-member.}
\tablecomments{This table is available in its entirety in the electronic edition of the journal.  A portion is reproduced here to provide guidance on form and content.}
\end{deluxetable*}

\subsection{Binarity}
\label{sec:binaries}

For member stars that were observed on at least two separate observing
runs, we check for velocity variations between measurements using a
$\chi^{2}$ test.  In most cases the data are consistent with the null
hypothesis that the velocity of the star is constant with time.  We
obtained multiple measurements of 15 of the 21 members of Gru~II.  Two
out of those 15 stars have probabilities of a constant velocity low
enough ($p \ll 0.01$) for that hypothesis to be confidently rejected:
the red giant DES~J220352.01$-$462446.5 ($\Delta V = 10.0 \pm
1.7$~\kms) and the blue horizontal branch (BHB) star
DES~J220433.75$-$462639.8 ($\Delta V = 14.5 \pm
3.3$~\kms).\footnote{For the latter star this velocity difference is
  measured between two spectra obtained a few hours apart during the
  same observing run.  It is therefore not clear whether to interpret
  the velocity measurements as evidence of a short-period binary
  system or evidence that one of the measurements is erroneous.  An
  additional velocity measurement for this star obtained $\sim1$~yr
  later agrees with the first measurement.}  Several other stars
exhibit weaker evidence for variability, with $0.01 \le p \le 0.10$.

For eight of the 11 members of Tuc~IV we obtained multiple spectra
separated by at least $\sim1$~month.  Two of these stars exhibit clear
velocity changes during the course of our observations
(DES~J000228.19$-$604814.3, $\Delta V = 15.7 \pm 3.1$~\kms;
DES~J000119.59$-$604439.2, $\Delta V = 13.0 \pm 2.8$~\kms), and a
third star with $p=0.03$ (DES~J000303.55$-$605025.3) may be variable
as well.  The first of these is a BHB star likely with an eccentric
orbit, as we observe slow velocity changes over several months in 2015
and a change of $\sim10$~\kms\ over less than 4 weeks in 2018.  The
other five members show no significant evidence for radial velocity
variability.  Finding two clear binaries out of seven stars in Tuc~IV
is a larger percentage of radial velocity variables than is seen in
other dwarf galaxies, although of course the binomial uncertainty is
large with so few stars.  Nevertheless, this result could suggest that
the binary fraction of Tuc~IV is very high, consistent with recent
determinations of the close binary fraction as a function of
metallicity in the Milky Way \citep{moe19} and with estimates of the
binary population in several satellite galaxies
\citep{geha13,spencer18,minor19}.  We observed two of the three Tuc~V
members with a time baseline of $\sim1$~yr, with no evidence for
velocity changes for either star.

\section{RESULTS}
\label{results}

\subsection{Structural Properties}
\label{sec:structure}

The luminosity, size, and other photometric parameters of our three
target satellites were originally determined from the DES year 2 quick
release catalog \citep{drlica15}, and have not been updated with
deeper data by other authors.\footnote{\citet{conn18a} did obtain deep
  Gemini imaging of Tuc~V, but were not able to characterize it (see
  below and \S~\ref{sec:discussion}).}  Because Tuc~IV and Tuc~V are
close to the detection limits of DES in surface brightness and
luminosity, respectively, we repeat the \citet{drlica15} analysis of
all three systems with the deeper and more uniform DES Y3A2 catalog
\citep{desdr1}.  The updated parameters for each object are listed in
Tables~\ref{gru2_table}--\ref{tuc5_table}.  Relative to the
\citeauthor{drlica15} results, the luminosity of all three satellites
is lower by $\gtrsim30\%$, although the error bars of the Y2 and Y3
luminosities overlap.  For Tuc~IV we find a somewhat smaller
half-light radius of 9\farcm3.  For Tuc~V we find a larger half-light
radius of 2\farcm1.  The larger size of Tuc~V may explain why
\citet{conn18a} were unable to identify a clear overdensity in the
small Gemini/GMOS field of view they employed.

\subsection{Stellar Kinematics}
\label{sec:kin}

Given the member samples identified in Section~\ref{sec:membership},
we use the maximum likelihood approach defined in \citet{simon17} and
\citet{li17} to determine the mean velocity and velocity dispersion of
each of the three satellites.  In cases where we have multiple
measurements of a star separated by more than $\sim1$ month, we
average all of the measurements together before computing the global
properties.

\begin{figure*}[th!]
\epsscale{1.15}
\plottwo{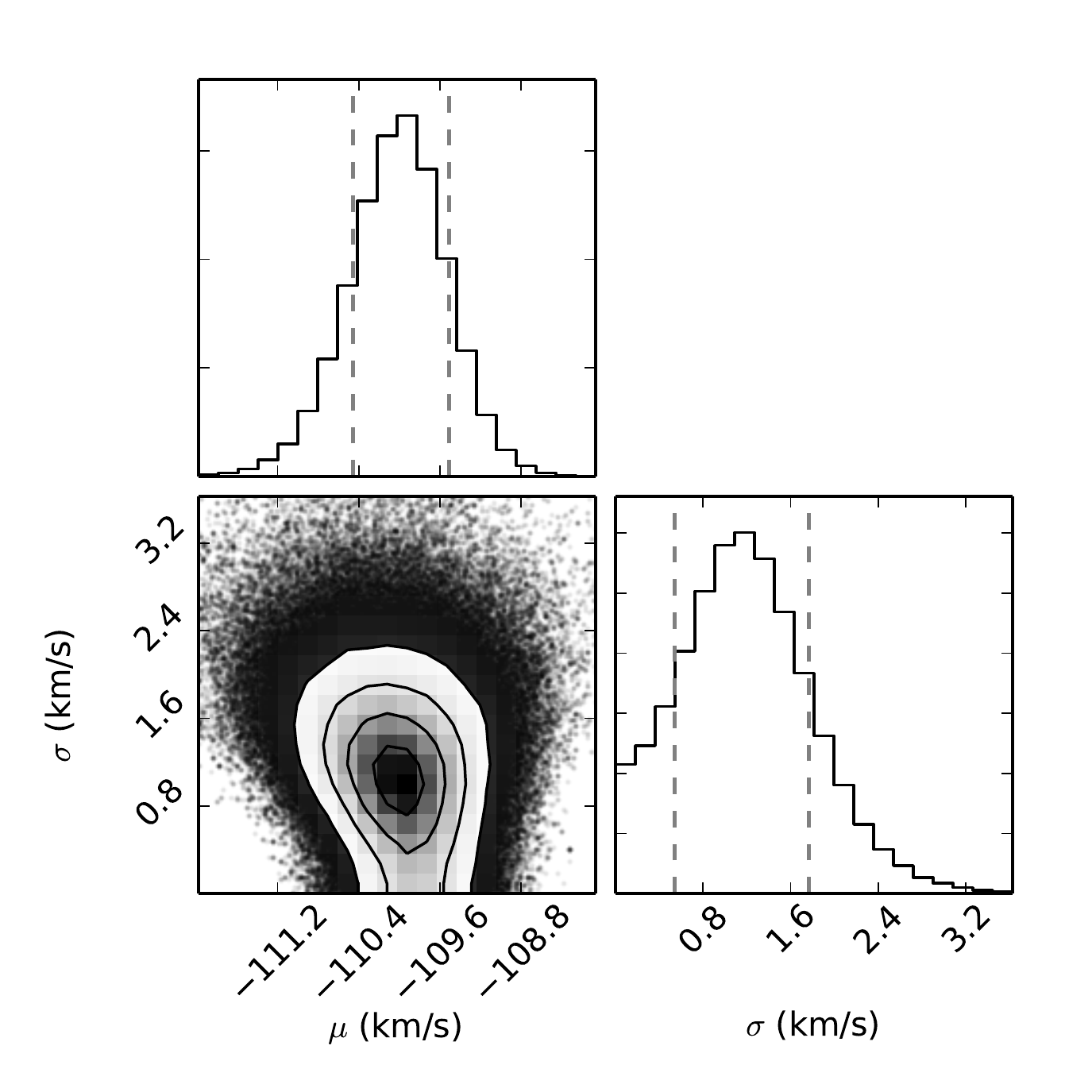}{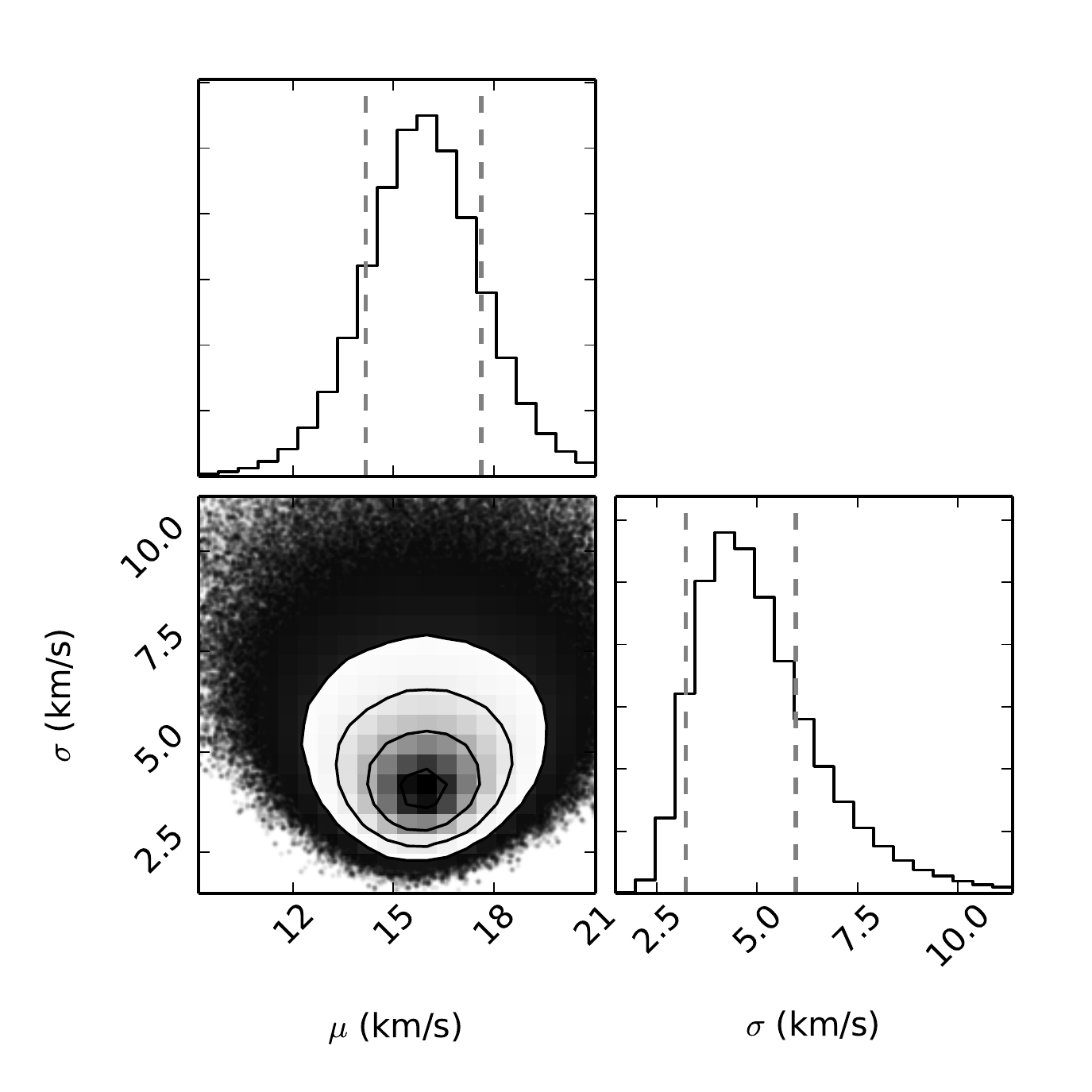}
\caption{(left) Corner plot for the velocity and velocity dispersion
  of Gru~II.  The upper limit on the velocity dispersion is 2.0~\kms.
  (right) Corner plot for the velocity and velocity dispersion of
  Tuc~IV.}
\label{vel_gru2}
\end{figure*}

For Gru~II, with the two binaries removed we measure a systemic
velocity of $v_{\rm hel} = -110.0 \pm 0.5$~\kms\ and a velocity
dispersion of $\sigma = 1.2 \pm 0.6$~\kms\ ($1\sigma$
uncertainties).  However, at $\sim2\sigma$ the dispersion is
consistent with zero, so our data do not clearly resolve the internal
kinematics of Gru~II.  We place a 90\%\ (95.5\%) confidence upper
limit on the velocity dispersion of 1.9~\kms\ (2.0~\kms).  The
posterior probability distributions for the velocity and velocity
dispersion are estimated using an affine invariant Markov Chain Monte
Carlo ensemble sampler \citep{fm13} and are displayed on the left side
of Figure~\ref{vel_gru2}.  Using the relation given by \citet{wolf10}
and the upper limit on the velocity dispersion of Gru~II, the
corresponding (90\%\ confidence) upper limit on the mass enclosed
within its half-light radius is $3.2 \times 10^{5}$~\msun\ (${\rm M/L}
< 300$~\msun/\lsun within the half-light radius).


\begin{deluxetable}{llr}
\tablecaption{Summary of Properties of Grus\,II}
\tablewidth{0pt}
\tablehead{
\colhead{Row} & \colhead{Quantity} & \colhead{Value}
}
\startdata
(1) & RA (J2000)                           & $331.025^{+0.009}_{-0.008}$ \\
(2) & Dec (J2000)                          & $-46.442 \pm 0.006$ \\
(3) & Distance (kpc)\tablenotemark{a}      & $55 \pm 2$  \\
(4) & $r_{\rm 1/2}$ (arcmin)\tablenotemark{b} & $5.9 \pm 0.5$ \\
(5) & Ellipticity                          & $<0.21$ \\
(6) & Position angle (degrees)             & ... \\
(7) & $M_{V,0}$                             & $-3.5 \pm 0.3$ \\
(8) & $L_{V,0}$ (L$_{\odot}$)               & $2100^{+700}_{-500}$ \\
(9) & $r_{\rm 1/2}$ (pc)                    & $94 \pm 9$  \\ 
(10)  & N$_{\rm spectroscopic~members}$         & 21 \\
(11)  & $V_{\rm hel}$ (\kms)                  & $-110.0 \pm 0.5$ \\
(12)  & $V_{\rm GSR}$ (\kms)                  & $-132.0 \pm 0.5$ \\
(13)  & $\sigma$ (\kms)\tablenotemark{c}      & $<2.0$ \\
(14)  & Mass (M$_{\odot}$)\tablenotemark{c}  & $<3.5 \times 10^{5}$  \\
(15)  & M/L$_{V}$ (M$_{\odot}$/L$_{\odot}$)\tablenotemark{c}  & $<330$ \\
(16)  & Mean [Fe/H]                        & $-2.51 \pm 0.11$ \\
(17)  & Metallicity dispersion (dex)\tablenotemark{c}       & $<0.45$ \\
(18)  & $\mua$ (\masyr)                    & $0.45 \pm 0.08$ \\
(19)  & $\mud$ (\masyr)                    & $-1.46 \pm 0.09$ \\
(20)  & Orbital pericenter (kpc)              & $25^{+6}_{-7}$ \\
(21)  & Orbital apocenter (kpc)               & $66^{+7}_{-5}$ \\
(22)  & $\log_{10}{J(0.2\degr)}$ (GeV$^{2}$~cm$^{-5}$)\tablenotemark{c} & $<16.5$ \\
(23)  & $\log_{10}{J(0.5\degr)}$ (GeV$^{2}$~cm$^{-5}$)\tablenotemark{c} & $<16.7$ \\

\enddata

  \tablenotetext{a}{We adopt the RR~Lyrae distance to Grus~II from
    \citet{martinez19}.  The RR Lyrae-based measurement is consistent
    with the isochrone distance but has a smaller uncertainty.}
  \tablenotetext{b}{The radius listed here is the semi-major axis of
    the half-light ellipse (referred to as $a_{\rm h}$ in \citealt{drlica15}).}
  \tablenotetext{c}{Upper limits listed here are at 95.5\%\ confidence.  See the text for values at other confidence levels.}
\label{gru2_table}
\end{deluxetable}

After excluding binary stars\footnote{Including the two binaries
  actually slightly \emph{decreases} the measured dispersion, at least
  in part because one of them has enough velocity measurements that
  the average of them is a reasonable estimate of the center of mass
  velocity of the binary.}, we measure a systemic velocity of $v_{\rm
  hel} = 15.9^{+1.8}_{-1.7}$~\kms\ and a velocity dispersion of $\sigma =
4.3^{+1.7}_{-1.0}$~\kms\ for Tuc~IV.  Despite the small sample of
member stars the dispersion of Tuc~IV is significantly resolved; at
$2\sigma$ the velocity dispersion is larger than 2.7~\kms.  The Monte
Carlo posteriors are displayed on the right side of
Figure~\ref{vel_gru2}.  Because the velocity of Tuc~IV coincides with
a substantial fraction of the Milky Way foreground stars, one might
worry that its velocity dispersion is being inflated by contamination
of the member sample.  However, since the stars classified as members
are all either BHB stars, where contamination should be negligible
(see Fig.~\ref{cmd_tuc4}a), or are RGB stars with both metallicities
and proper motions consistent with membership, such contaminants are
unlikely.  The nonzero velocity dispersion in Tuc~IV is based
primarily on the velocities of three stars (DES~J000233.86$-$605439.4,
DES~J000251.92$-$604820.8, and DES~J000344.06$-$604804.7), the first
two of which have velocities $\sim1.5\sigma$ below the mean velocity
of the system, and the third $\sim1.5\sigma$ above the mean velocity.
If (some of) these stars are binaries, the actual intrinsic dispersion
of Tuc~IV could be smaller.  However, we have multiple velocity
measurements of each star spanning from 3 months
(DES~J000344.06$-$604804.7) to 3 years for the other two, with no
evidence of radial velocity variations, so we can rule out that they
have binary companions with periods shorter than a few years.

\begin{deluxetable}{llr} [t!]
\tablecaption{Summary of Properties of Tucana\,IV}
\tablewidth{0pt}
\tablehead{
\colhead{Row} & \colhead{Quantity} & \colhead{Value}
}
\startdata
(1) & RA (J2000)                           & $0.717^{+0.014}_{-0.021}$ \\
(2) & Dec (J2000)                          & $-60.830^{+0.010}_{-0.011}$ \\
(3) & Distance (kpc)                       & $47 \pm 4$  \\
(4) & $r_{\rm 1/2}$ (arcmin)\tablenotemark{a} & $9.3^{+1.4}_{-0.9}$ \\
(5) & Ellipticity                          & $0.39^{+0.07}_{-0.10}$ \\
(6) & Position angle (degrees)             & $27^{+9}_{-8}$ \\
(7) & $M_{V,0}$                             & $-3.0^{+0.3}_{-0.4}$ \\
(8) & $L_{V,0}$ (L$_{\odot}$)               & $1400^{+600}_{-300}$ \\
(9) & $r_{\rm 1/2}$ (pc)                    & $127^{+22}_{-16}$  \\ 
(10)  & N$_{\rm spectroscopic~members}$         & 11 \\
(11)  & $V_{\rm hel}$ (\kms)                  & $15.9^{+1.8}_{-1.7}$ \\
(12)  & $V_{\rm GSR}$ (\kms)                  & $-82.9^{+1.8}_{-1.7}$ \\
(13)  & $\sigma$ (\kms)                     & $4.3^{+1.7}_{-1.0}$ \\
(14)  & Mass (M$_{\odot}$)                  & $2.2^{+1.8}_{-1.1} \times 10^{6}$  \\
(15)  & M/L$_{V}$ (M$_{\odot}$/L$_{\odot}$)  & $3100^{+2900}_{-1600}$ \\
(16)  & Mean [Fe/H]                        & $-2.49^{+0.15}_{-0.16}$ \\
(17)  & Metallicity dispersion (dex)\tablenotemark{b}       & $<0.64$ \\
(18)  & $\mua$ (\masyr)                    & $0.51 \pm 0.14$ \\
(19)  & $\mud$ (\masyr)                    & $-1.64 \pm 0.13$ \\
(20)  & Orbital pericenter (kpc)              & $25^{+11}_{-8}$ \\
(21)  & Orbital apocenter (kpc)               & $52^{+12}_{-6}$ \\
(22)  & $\log_{10}{J(0.2\degr)}$ (GeV$^{2}$~cm$^{-5}$) & $18.2^{+0.6}_{-0.5}$ \\
(23)  & $\log_{10}{J(0.5\degr)}$ (GeV$^{2}$~cm$^{-5}$) & $18.4^{+0.6}_{-0.5}$ \\

\enddata

  \tablenotetext{a}{The radius listed here is the semi-major axis of
    the half-light ellipse (referred to as $a_{\rm h}$ in \citealt{drlica15}).}
  \tablenotetext{b}{Upper limits listed here are at 95.5\%\ confidence.  See the text for values at other confidence levels.}
\label{tuc4_table}
\end{deluxetable}

The mass within the half-light radius of Tuc~IV according to the
\citet{wolf10} estimator is $2.2^{+1.8}_{-1.1} \times 10^{6}$~\msun.
However, we note that the uncertainty on the mass is non-Gaussian, so
the $2\sigma$ uncertainty is not twice as large as the $1\sigma$
uncertainty.  Even using the $2\sigma$ lower limit on the velocity
dispersion, the half-light mass of Tuc~IV is large ($8.6 \times
10^{5}$~\msun).  For comparison, the total luminosity of Tuc~IV is
1400~\lsun, indicating a highly dark matter-dominated system.

\begin{figure}[th!]
\epsscale{1.15}
\plotone{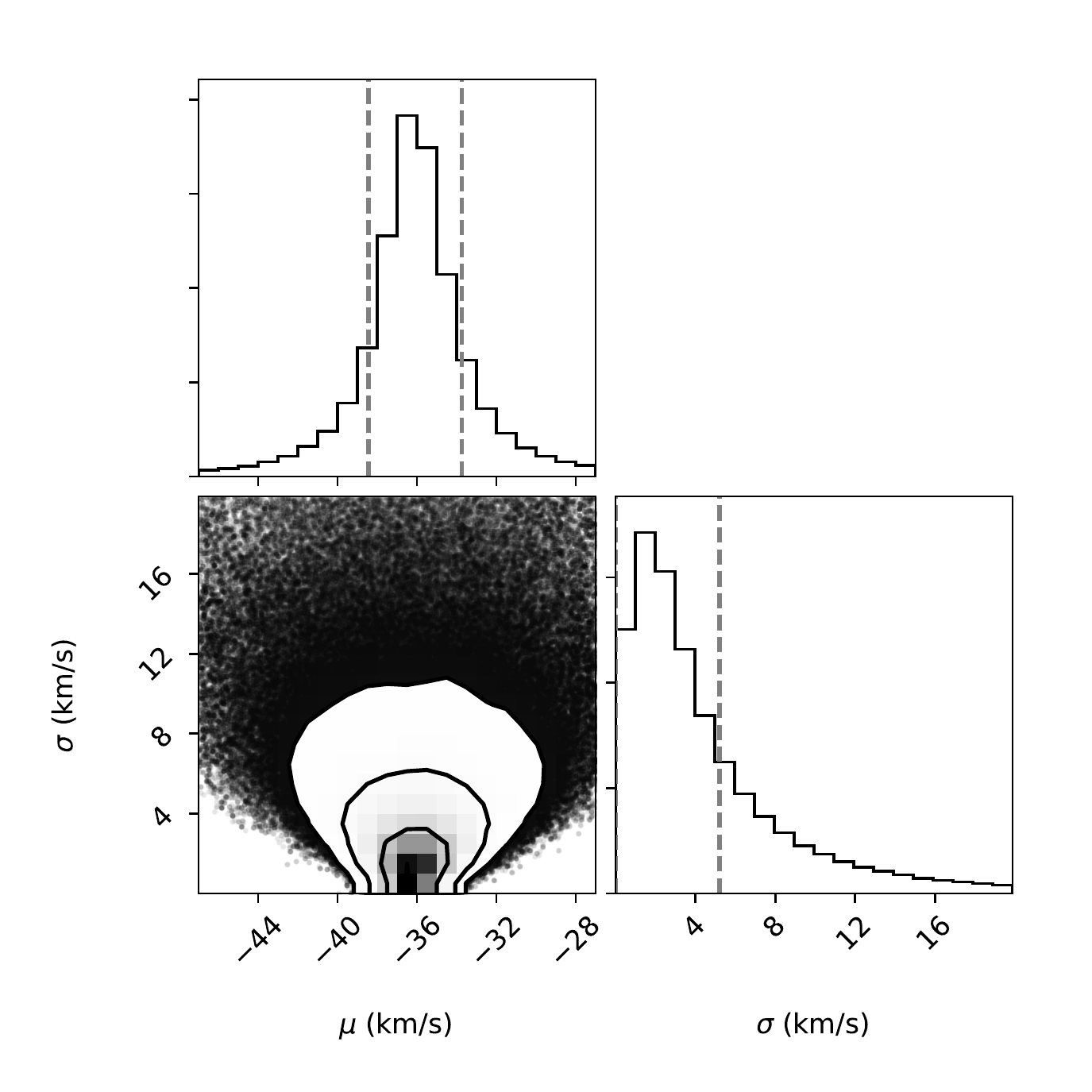}
\caption{Corner plot for the velocity and velocity dispersion of
  Tuc~V.}
\label{vel_tuc5}
\end{figure}

From the three members of Tuc~V we determine a systemic velocity of
$v_{\rm hel} = -36.2^{+2.5}_{-2.2}$~\kms.  Given the small sample and
the similar velocities of the three stars we cannot place any
significant constraints on the velocity dispersion of the system (the
$2\sigma$ upper limit is 7.4~\kms).  Substantially deeper spectroscopy
will be needed to study the internal kinematics of Tuc~V.  The
posterior probability distributions are displayed in
Figure~\ref{vel_tuc5}.  The mass of Tuc~V is not usefully constrained
by the data.
  
\begin{deluxetable}{llr}
\tablecaption{Summary of Properties of Tucana\,V}
\tablewidth{0pt}
\tablehead{
\colhead{Row} & \colhead{Quantity} & \colhead{Value}
}
\startdata
(1) & RA (J2000)                           & $354.347 \pm 0.008$ \\
(2) & Dec (J2000)                          & $-63.266^{+0.006}_{-0.004}$ \\
(3) & Distance (kpc)                       & $55^{+3}_{-8}$  \\
(4) & $r_{\rm 1/2}$ (arcmin)\tablenotemark{a} & $2.1^{+0.6}_{-0.4}$ \\
(5) & Ellipticity                          & $0.51^{+0.09}_{-0.18}$ \\
(6) & Position angle (degrees)             & $29 \pm 11$ \\
(7) & $M_{V,0}$                             & $-1.1^{+0.5}_{-0.6}$ \\
(8) & $L_{V,0}$ (L$_{\odot}$)               & $240^{+170}_{-90}$ \\
(9) & $r_{\rm 1/2}$ (pc)                    & $34^{+11}_{-8}$  \\ 
(10)  & N$_{\rm spectroscopic~members}$         & 3 \\
(11)  & $V_{\rm hel}$ (\kms)                  & $-36.2^{+2.5}_{-2.2}$ \\
(12)  & $V_{\rm GSR}$ (\kms)                  & $-136.6^{+2.5}_{-2.2}$ \\
(13)  & $\sigma$ (\kms)\tablenotemark{b}      & $<7.4$ \\
(14)  & Mass (M$_{\odot}$)\tablenotemark{b}  & $<1.7 \times 10^{6}$  \\
(15)  & M/L$_{V}$ (M$_{\odot}$/L$_{\odot}$)\tablenotemark{b}  & $<14000$ \\
(16)  & Mean [Fe/H]                        & $-2.17 \pm 0.23$ \\
(17)  & $\mua$ (\masyr)                    & $-0.62 \pm 0.31$ \\
(18)  & $\mud$ (\masyr)                    & $-0.88 \pm 0.35$ \\
(19)  & Orbital pericenter (kpc)              & $36^{+13}_{-15}$ \\
(20)  & Orbital apocenter (kpc)               & $131^{+470}_{-60}$ \\
(21)  & $\log_{10}{J(0.2\degr)}$ (GeV$^{2}$~cm$^{-5}$)\tablenotemark{b} & $<19.4$ \\
(22)  & $\log_{10}{J(0.5\degr)}$ (GeV$^{2}$~cm$^{-5}$)\tablenotemark{b} & $<19.6$ \\

\enddata

  \tablenotetext{a}{The radius listed here is the semi-major axis of
    the half-light ellipse (referred to as $a_{\rm h}$ in \citealt{drlica15}).}
  \tablenotetext{b}{Upper limits listed here are at 95.5\%\ confidence.}
\label{tuc5_table}
\end{deluxetable}

\subsection{Metallicities and Metallicity Spreads}
\label{sec:metallicity}

We use the same maximum likelihood method to determine the metallicity
distribution of each satellite.  The RGB stars in Gru~II have a mean
metallicity of $\feh = -2.51 \pm 0.11$, with a dispersion of
$\sigma_{\rm [Fe/H]} = 0.21^{+0.15}_{-0.13}$~dex.  The $2\sigma$ upper
limit on the metallicity spread is 0.45~dex.  The mean metallicity of
Tuc~IV is nearly identical, at $\feh = -2.49^{+0.15}_{-0.16}$.  The
metallicity dispersion for the eight RGB stars in Tuc~IV is
$\sigma_{\rm [Fe/H]} = 0.18^{+0.20}_{-0.18}$~dex, with a $2\sigma$
upper limit of 0.64~dex.  The measurements for Tuc~V are also similar,
with $\feh = -2.17 \pm 0.23$ and a metallicity dispersion that cannot
be significantly constrained with only two metallicity measurements
(the S/N for the third Tuc~V member star is too low for a reliable
determination of the CaT EW).

\subsection{Proper Motions and Orbits}

Given a set of spectroscopic members in each object, we can use the
astrometry provided by the second data release \citep{gaiadr2brown}
from the Gaia mission \citep{gaia16a} to measure mean proper motions,
as in \citet{simon18}.  \citet{mh18} and \citet{pl19} already
estimated the proper motions of Gru~II and Tuc~IV using
photometrically-selected member samples, but our spectroscopic samples
offer a cleaner and more robust determination.  Using only
spectroscopic members, we measure proper motions of $\mua = 0.45 \pm
0.08$~\masyr, $\mud = -1.46 \pm 0.09$~\masyr\ for Gru~II and $\mua =
0.51 \pm 0.14$~\masyr, $\mud = -1.64 \pm 0.13$~\masyr\ for Tuc~IV.
These values are in excellent agreement with the determinations of
\citet{pl19}, as well as with the \citet{mh18} measurements in the
Dec. direction, but deviate by $\sim1.5\sigma$ from \citet{mh18} in
the R.A. direction.  We find that up to $\sim1/4$ of the Gru~II stars
and $\sim1/3$ of the Tuc~IV stars assumed to be members by
\citet{mh18} are not spectroscopic members, which may explain this
discrepancy.  For Tuc~V, which has no previous proper motion
measurement, we derive $\mua = -0.62 \pm 0.31$~\masyr, $\mud =
-0.88 \pm 0.35$~\masyr\ from the three spectroscopic members.

\subsubsection{Orbits in the Milky Way Potential} \label{sec:orbits}

In combination with the radial velocities measured here
(\S~\ref{sec:kin}) and the distances listed in
Tables~\ref{gru2_table}-\ref{tuc5_table}, these proper motions
determine the orbits of the three satellites around the Milky Way.  As
a starting point, we adopt the modified MWPotential2014 (where the
original MWPotential2014 is taken from \citealt{galpy}) gravitational
potential defined by \citet{carlin18}, with a total mass for the Milky
Way of $1.6 \times 10^{12}$~\msun \citep[e.g.,][]{watkins19}.  We
integrate orbits with \texttt{galpy} \citep{galpy} as described in
\citet{simon18}.  The orbit of Gru~II is tightly bound and
well-determined, with a pericenter of $25^{+6}_{-7}$~kpc (where the
confidence interval is determined from 1000 Monte Carlo iterations,
drawing input distances, radial velocities, and proper motions from
distributions set by the measured values and their uncertainties), an
apocenter of $66^{+7}_{-5}$~kpc, an eccentricity of
$0.45^{+0.08}_{-0.05}$, and an orbital period of $\sim0.9~$~Gyr.
Gru~II has likely completed many orbits around the Galaxy, and it is
currently approaching pericenter, which it will reach in
$\sim200$~Myr.  The situation for Tuc~IV is similar, with a best-fit
orbital pericenter of $25^{+11}_{-8}$~kpc, an apocenter of
$52^{+12}_{-6}$~kpc, an eccentricity of $0.36^{+0.13}_{-0.06}$, and an
orbital period of $\sim0.7~$~Gyr.
Like Gru~II, Tuc~IV is approaching the pericenter of its orbit and
will reach that point in $\sim200$~Myr.  These two satellites lessen
the previously observed tendency of known ultra-faint dwarfs to be
discovered near their pericenters \citep{simon18,fritz18}.

Tuc~V has a somewhat larger orbital pericenter than the other two
systems, $35^{+10}_{-13}$~kpc, and its best-fit apocenter reaches well
out into the halo of the Milky Way ($d_{\rm apo} =
126^{+243}_{-49}$~kpc).  The corresponding orbital eccentricity is
$0.59^{+0.21}_{-0.07}$.  Tuc~V has a median orbital period of 1.3~Gyr,
but in $\sim20$\% of the Monte Carlo iterations Tuc~V is approaching
the Milky Way for the first time.  Its next (or first) pericenter will
occur in $\sim145$~Myr.

In order to test the robustness of these orbital results, we also
compute the orbits of the three satellites in the \cite{mcmillan17}
Milky Way potential using \textsc{galpot} \citep{db98}. We find that
most of the orbital parameters are within the $1\sigma$ ranges given
in Tables~\ref{gru2_table}, \ref{tuc4_table}, and \ref{tuc5_table}.

\subsubsection{A Collision with the LMC}
\label{sec:lmc_collision}

\begin{figure*}[t!]
\includegraphics[width=6cm]{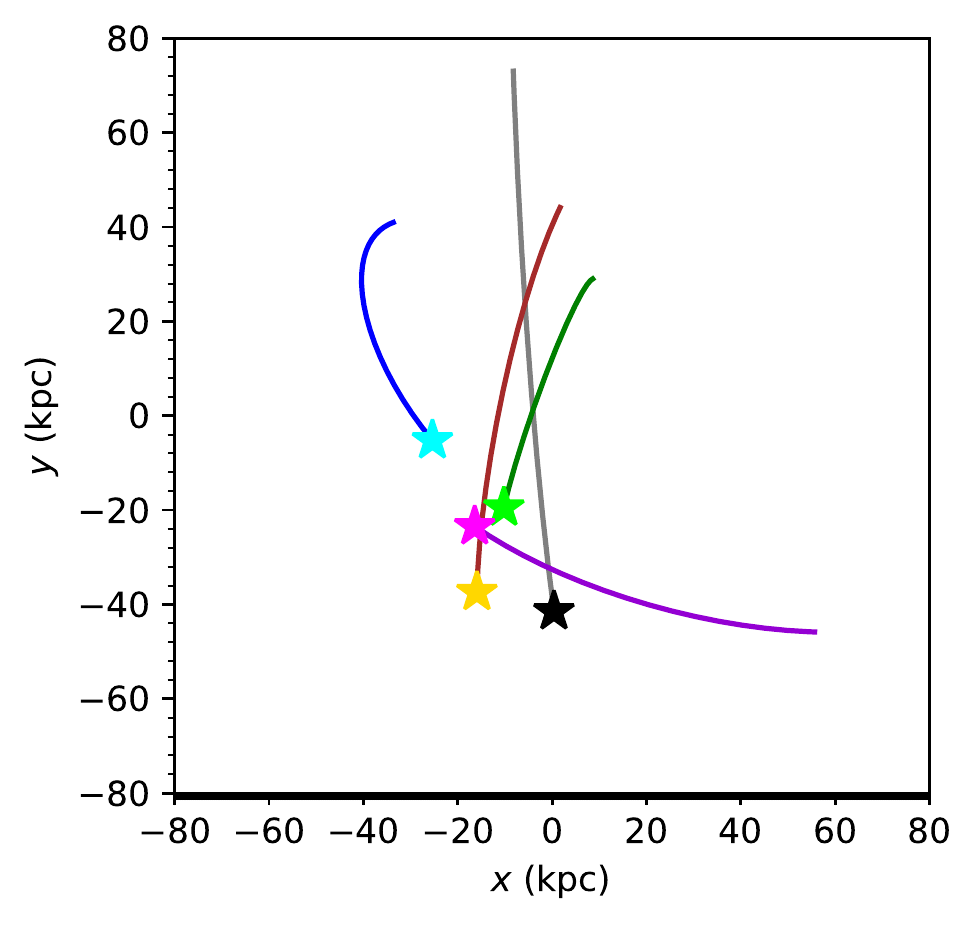}\includegraphics[width=6cm]{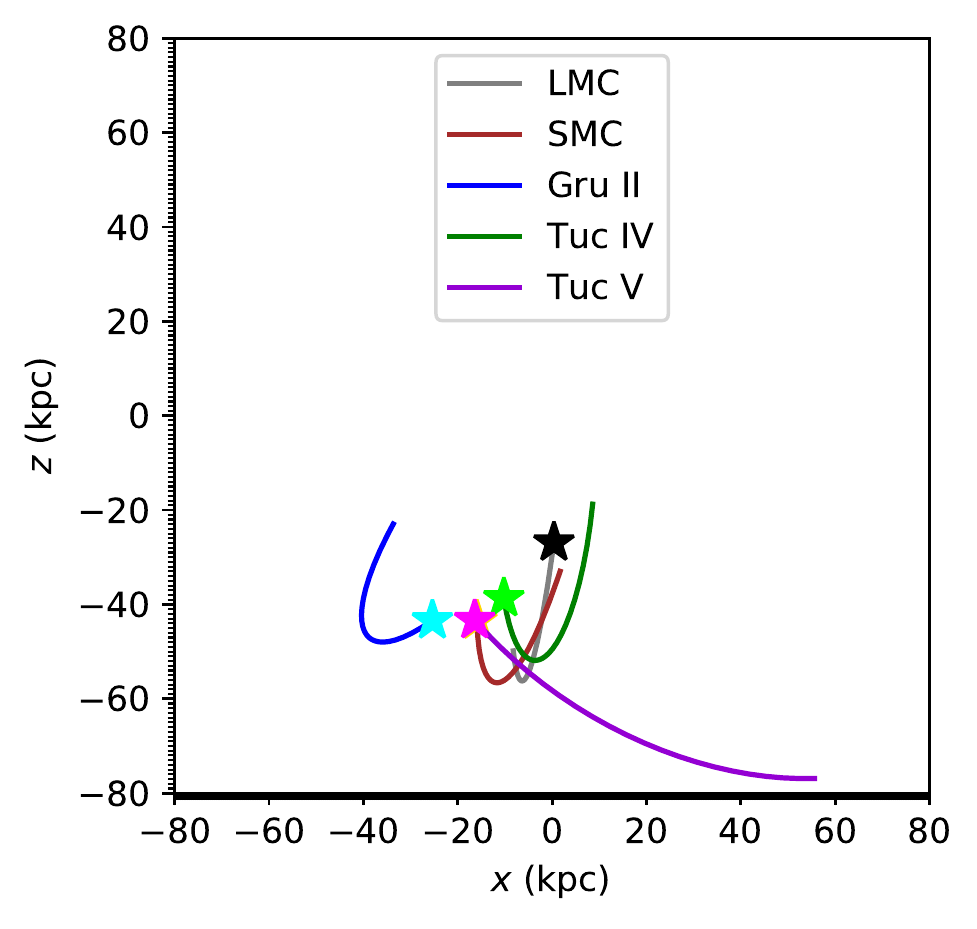}\includegraphics[width=6cm]{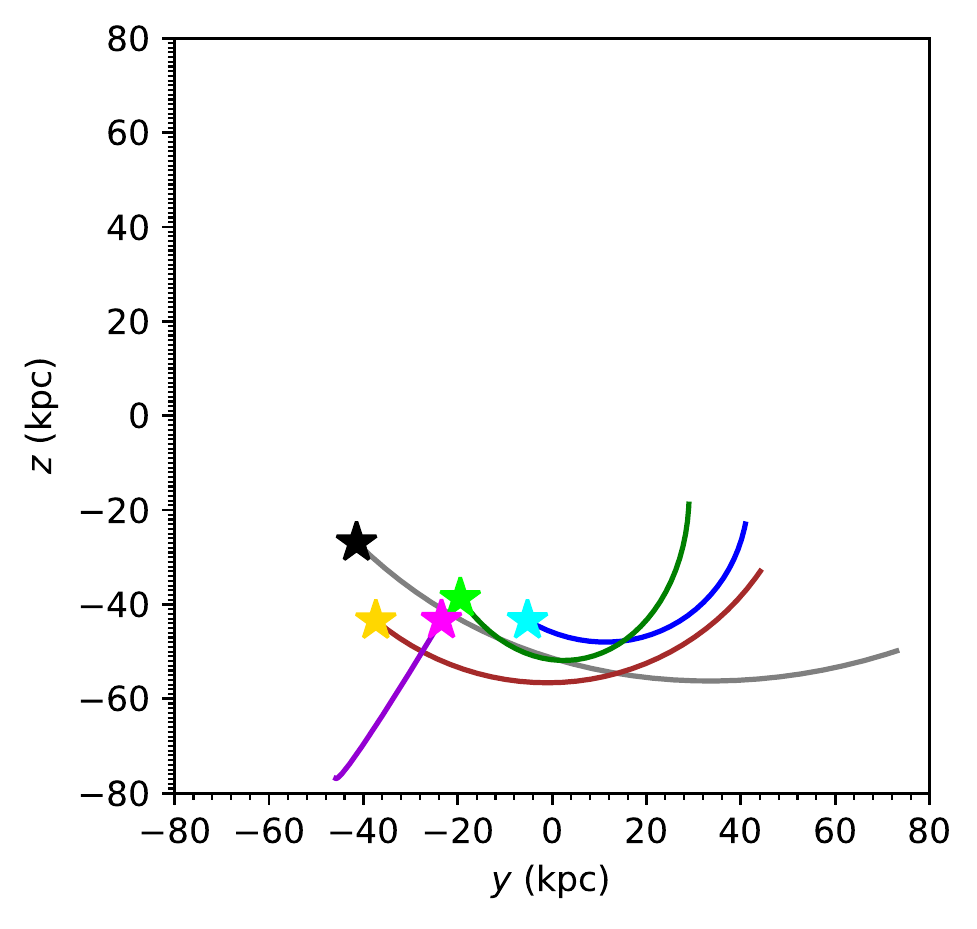}
\caption{(left) Projections of the orbits on the X-Y plane for Gru~II
  (blue), Tuc~IV (green), and Tuc~V (purple), as well as the LMC
  (gray) and SMC (brown).  The orbits are integrated backward for
  400~Myr, and the star at the end of each trajectory indicates the
  present position of the object.  (middle) Projections of the orbits
  on the X-Z plane for the same dwarfs. (right) Projections of the
  orbits on the Y-Z plane.  The very different paths by which the
  three ultra-faint satellites have reached the current apparent
  grouping are clear.  The LMC and Tuc~IV orbits intersect
  $\sim150$~Myr ago in all three planes.  }
\label{orbits_lmc}
\end{figure*}

To assess whether any of the three satellites could be associated with
the Magellanic Clouds, we integrate the derived orbits and that of the
LMC backward in the Milky Way potential, as shown in
Figure~\ref{orbits_lmc}.  For the LMC we adopt the proper motion from
\citet{kallivayalil13}, the radial velocity from \citet{vdm02}, and we
assume a distance of $49.97\pm1.13$~kpc from \cite{pietrzynski13}.
Gru~II and Tuc~V follow very different paths from the LMC and are
unlikely to be associated with it (see \S~\ref{mc_assoc} for more
detailed discussion).  We find that Tuc~IV, on the other hand, passed
$5.8$~kpc from the center of the LMC with a relative velocity of
$\sim200$ \kms $\sim130$~Myr ago.  This close encounter may have both
deflected Tuc~IV from its previous trajectory and disturbed its
internal kinematics, and renders our simple calculation ignoring the
gravitational potential of the LMC incomplete. For comparison,
although Tuc~V also had a relatively close approach to the LMC of
$16.5$ kpc $\sim 50$ Myr ago, the relative velocity between the two
was $\sim460$ \kms, suggesting that it is not a Magellanic
satellite. For the rest of this section we will only focus on Tuc~IV
since Gru~II and Tuc~V are not strongly affected by the LMC.

In order to include the effect of the Magellanic Clouds, we repeat the
orbit calculations following a similar approach to those described by
\citet{erkal18,erkal19} and \citet{eb19}.  Motivated by the
measurement of the LMC mass in \cite{erkal19}, we model the LMC as a
Hernquist profile \citep{hernquist90} with a mass of
$1.5\times10^{11}$~M$_\odot$ and a scale radius of 17.13~kpc. We use
the same potential for the Milky Way as in Section \ref{sec:orbits}
and account for the reflex motion of the Milky Way in response to the
LMC as in \cite{erkal19}. We find that Tuc~IV collided with the LMC
with a closest approach of $4.1^{+3.2}_{-2.2}$~kpc and a relative
velocity of $281^{+26}_{-39}$~\kms at a time of $119^{+26}_{-18}$~Myr
ago.

As in \citet{eb19}, this rewinding procedure can be used to determine
whether Tuc~IV was originally an LMC satellite by computing the
likelihood that Tuc~IV was bound to the LMC prior to their
encounter. The resulting low probability of 3.8\% suggests that it is
not a Magellanic satellite. However, given the close passage of Tuc~IV
with respect to the LMC, it is natural to include the SMC as well. For
the SMC we use the proper motion from \citet{kallivayalil13}, the
radial velocity from \cite{hz06}, and the distance from
\cite{graczyk14}.  If we treat the SMC as a Hernquist profile with a
mass of $10^{10}$~M$_\odot$ and a scale radius of 1~kpc, the
probability that Tuc~IV was originally bound to the LMC rises to
18.1\%. While this probability is still modest compared to the
satellites classified as having a Magellanic origin by \citet{eb19}
($\gtrsim 50\%$), we have re-run the analysis of \citeauthor{eb19}
including the SMC and we find that Tuc~IV is the only satellite for
which including the SMC has an appreciable effect. The closest
approach between Tuc~IV and the SMC is $6.6^{+5.3}_{-3.7}$ kpc, which
is comparable to the closest approach with the LMC. This result
suggests that Tuc~IV may have actually undergone a three-body
interaction with both Magellanic Clouds, which can be tested once
improved proper motions are available.

Given the close passage of Tuc~IV with respect to the LMC, we note
that although the Hernquist profile we have assumed for the LMC
matches the observed rotation curve from \cite{vdmk14} at 8.7~kpc, the
modeled rotation curve falls below the observed one at smaller
radii. Thus, we may be underpredicting the effect of the LMC on
Tuc~IV. In order to assess the importance of this discrepancy, we also
consider an LMC whose gravitational field matches the previous
Hernquist profile beyond 8.7~kpc but has a flat rotation curve at
91.7~\kms within 8.7~kpc, in approximate agreement with a variety of
kinematic data for the LMC.  We find that this change only affects the
probability of Tuc IV being an LMC satellite at the $\sim1\%$ level
and thus does not influence our conclusions.

\begin{figure}[th!]
\epsscale{1.15}
\plotone{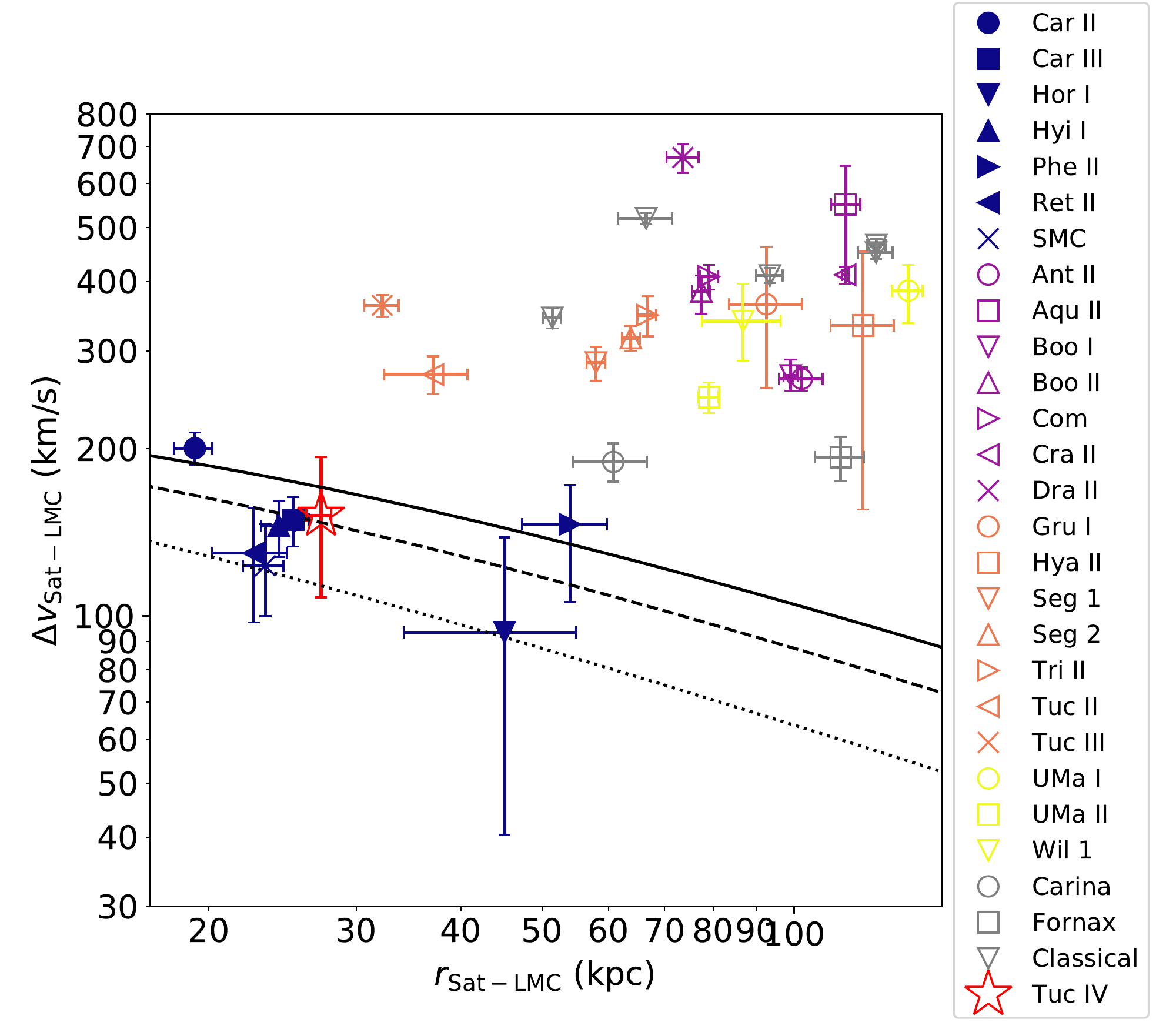}
\caption{Distance and speed of satellites with respect to the
  LMC. This figure is almost identical to the one in
  \protect\cite{eb19} except we have also included the relative
  distance and velocity of Tuc~IV. The dotted, dashed, and solid lines
  show the escape velocity for an LMC with masses of
  $5,10,15\times10^{10}$~M$_\odot$ and scale radii of $6.2, 12.4,
  17.1$~kpc, respectively.  The solid blue markers show the satellites
  that \protect\cite{eb19} identified as Magellanic
  satellites. Interestingly, Tuc~IV sits at a similar position in
  phase-space.}
\label{fig:energy_tuc5}
\end{figure}

Interestingly, even though we have assigned Tuc~IV a modest
probability of being a Magellanic satellite, its present-day position
and velocity relative to the LMC are comparable to those of the
Magellanic satellites in \cite{eb19}. Figure \ref{fig:energy_tuc5}
shows the position and velocity relative to the LMC for the satellites
considered in \cite{eb19} and Tuc~IV. Tuc~IV sits in a similar
location in this space as the Magellanic satellites.

Given that Tuc~IV has passed directly through the LMC in the recent
past, we next evaluate the tidal shock it experienced from the LMC.
At a radius of 4.1~kpc, the LMC has a circular velocity of
$\sim90$~\kms\ \citep{olsen11,vdmk14,vasiliev18}.  The corresponding
enclosed mass is $7.7 \times 10^{9}$~M$_{\odot}$.  Naively taking the
mass within the half-light radius of Tuc~IV from \S~\ref{sec:kin}, the
Jacobi radius of Tuc~IV at the time of minimum distance to the LMC
would be 188~pc, $\sim1.5$ times its present half-light radius.  If
much of the dark matter halo of Tuc~IV had previously been stripped by
encounters with either the Magellanic Clouds or the Milky Way, we
would expect this close passage to significantly disturb its internal
structure.\footnote{It may also be interesting to consider whether the
  impact of Tuc IV left observable imprints on the disk of the LMC.}
In particular, if the true (undisturbed) velocity dispersion of Tuc~IV
is less than 2.4~\kms\ then the tidal radius of Tuc~IV during the
closest approach to the LMC would be within its half-light radius,
implying a major distortion that would likely affect the observed
kinematics.  However, if Tuc~IV retained a massive, extended dark
matter halo until this recent event then its tidal radius probably
remained at $\gtrsim 3 r_{1/2}$, insulating its stars from the tidal
influence of the LMC.  It is therefore possible in principle for
Tuc~IV to have survived this interaction without major damage to its
stellar component.  We note that several crossing times of Tuc~IV have
elapsed since the collision, so the system could have reached a new
equilibrium even if it was significantly perturbed by the encounter.
The uncertainties on the impact parameter, the inner mass distribution
of the LMC, and the dynamical mass of Tuc~IV are large enough that
stronger conclusions would require much more detailed calculations and
N-body simulations.

\section{DISCUSSION}
\label{sec:discussion}

One of the primary goals of this study is to determine the nature of
Gru~II, Tuc~IV, and Tuc~V.  Ideally, satellite classifications are
based on direct (a dynamical mass significantly larger than the
stellar mass) or indirect (retention of supernova ejecta) evidence for
the presence of dark matter \citep{ws12}.  However, as fainter and
fainter systems are studied, the detection of non-zero velocity and/or
metallicity dispersions becomes more and more difficult
\citep[e.g.,][]{martin16b,kirby17,simon17,longeard18,ji19}.  

Even without spectroscopy, the large physical sizes ($r_{\rm half} >
90$~pc) of Gru~II and Tuc~IV strongly suggest that they are galaxies.
While outer halo globular clusters can have half-light raii of
$\sim20$~pc, and in extreme cases tidal shocking of clusters at their
orbital pericenter can briefly lead to sizes as large as $\sim40$~pc
\citep[e.g.,][]{contenta17}, there are no known processes that will
inflate the actual or apparent radius of a cluster to $\sim100$~pc.
The observed internal kinematics of Tuc~IV are consistent with this
classification, with a mass-to-light ratio within its half-light
radius of $\sim3100$~M$_{\odot}$/L$_{\odot}$.  We therefore consider
Tuc~IV to be a spectroscopically confirmed dwarf galaxy.  For Gru~II,
we are not able to detect spectroscopically any evidence for dark
matter, although the measurements certainly do not rule out a
substantial dark matter content either.  The mean metallicity of
Gru~II is similar to or just below those of the most metal-poor known
globular clusters \citep[e.g.,][]{sobeck11,simpson18}, providing some
support for the idea that it is a dwarf.  Combining that information
with its size, we conclude that Gru~II is very likely a galaxy,
although more direct evidence would still be desirable.

For Tuc~V, we cannot rely on any of the above arguments since we lack
a resolved velocity or metallicity dispersion.  However, its revised
half-light radius of 34~pc (Section~\ref{sec:structure}) is now more
consistent with a dwarf galaxy classification, as it (at least
slightly) exceeds the size of known clusters.  The mean metallicity of
$\feh = -2.17$ is $\sim0.5$~dex higher than would be expected for a
dwarf of its luminosity according to the luminosity-metallicity
relation of \citet{kirby13b}, but since this measurement is based on
just 2 stars the significance of the discrepancy is not large.  Only a
small fraction of the Galactic globular cluster population is located
at distances comparable to that of Tuc~V, but among the clusters in
the outer halo the orbital properties of Tuc~V would not stand out
\citep[e.g.,][]{baumgardt19}.  Based on the available data, we regard
the nature of Tuc~V as undetermined until further spectroscopic
observations are obtained.

\citet{conn18a} obtained Gemini/GMOS-S imaging of Tuc~V extending
$\sim3$~mag deeper than the DES imaging in which Tuc~V was discovered.
They clearly detected the stellar population identified as Tuc~V by
\citet{drlica15}, but based on the apparent irregularity of the
spatial distribution of those stars they concluded that Tuc~V is
either a chance grouping of stars in the Small Magellanic Cloud (SMC),
a structure in the SMC halo, or a tidally disrupted star cluster
rather than a bound stellar system.  We disagree with their
interpretation of Tuc~V.  Our spectroscopy shows that the mean
velocity of Tuc~V is offset from that of the SMC by nearly 200~\kms,
ruling out the hypothesis that Tuc V is an overdensity (either real or
a line-of-sight projection) associated with the SMC.  As mentioned
above, the spectroscopic measurements cannot currently distinguish
between a dwarf galaxy and a globular cluster, but in either case
there is no significant evidence favoring tidal disruption.  Analyses
by \citet{martin08}, \citet{walsh08}, and \citet{munoz10} have shown
that irregular low surface brightness features around ultra-faint
satellites such as those seen in the vicinity of Tuc~V are generally
not statistically significant, and instead are consistent with being
the result of Poisson fluctuations from drawing a small number of
stars from a smooth spatial distribution.  Photometric uncertainties
and star-galaxy contamination at faint magnitudes can also contribute
to the apparent presence of tidal debris in ground-based imaging
\citep{mp19}.  Based on the revised size of Tuc~V determined in
Section~\ref{sec:structure}, we suggest that the small GMOS field of
view ($5\farcm5 \times 5\farcm5$) hinders a robust structural analysis
of the system.  In particular, \citet{munoz12} showed that accurately
determining the half-light radius of a low surface brightness stellar
system requires imaging over a field of view $>3\times$ the half-light
radius.  For the previously published size of Tuc~V GMOS meets this
criterion, but with our new measurement it does not.

\subsection{J-Factor}

The Milky Way's dwarf galaxies are among the best targets for searches
for dark matter annihilation and decay radiation because they have
high dark matter densities, are located close to the Sun, and are
nearly background free.  Analyses of $\gamma$-rays from the LAT
instrument on the Fermi telescope probe the thermal cross section
\citep{ackermann15, albert17}.  The calculation of the predicted dark
matter flux is split into two components. The first is dependent on
the distribution of dark matter within the dwarf (the astrophysics
component) and the second is related to properties of the dark matter
particle(s), such as the cross sections or mass (particle physics
component).  Adding new dwarf galaxies to the searches improves the
reach in dark matter parameter space.

The astrophysics components of the calculation mentioned above are
commonly referred to as the J-factor and D-factor for dark matter
annihilation and decay, respectively.  The J-factor is the integral
over the line-of-sight of the square of the dark matter density,
$J(\theta) =\int\rho_{\rm DM}^2 \mathrm{d}\Omega\mathrm{d}l$, and the
D-factor is the linear analog, $D(\theta)=\int\rho_{\rm DM}
\mathrm{d}\Omega\mathrm{d}l$.  The standard approach to measuring
$\rho_{\rm DM}$ is to solve the spherical Jeans equation using the
observed line-of-sight velocity dispersion \citep[e.g.,
][]{bonnivard15, strigari18}.

Briefly, we solve the spherical Jeans equations, project the velocity
dispersion into the line-of-sight direction, and compare the stellar
velocity data to the model predictions to determine dark matter
parameter distributions.  We assume the stellar distribution follows a
\citet{plummer11} profile, the dark matter density profile follows an
NFW profile \citep{nfw96}, and the stellar anisotropy is constant with
radius.  We treat the distance, ellipticity, and stellar half-light
radius as free parameters with Gaussian priors to account for their
measurement errors.  This framework is similar in scope to most
J-factor analyses for Milky Way dwarfs \citep[e.g., ][]{strigari08,
  bonnivard15}.  For additional details see \citet{ps19}.

Given the upper limit on the velocity dispersion for Gru~II, we are
similarly only able to set an upper limit on the J-factor (see
Table~\ref{gru2_table}).  We also find $\log_{10}{D}<16.6, 17.2$ at
solid angles of $0.2\degree,0.5\degree$.  For Tuc~IV we calculate
integrated J-factors of $\log_{10}{J}=18.0\pm0.6, 18.2^{+0.6}_{-0.5},
18.4^{+0.6}_{-0.5}$ within solid angles of $\theta=0.1\degree,
0.2\degree, 0.5\degree$ in logarithmic units of ${\rm GeV^{2} \,
  cm^{-5}}$ (i.e., $\log_{10}{(J(\theta)/{\rm GeV^{2} \, cm^{-5}})}$)
and integrated D-factors of $\log_{10}{D}=17.0\pm0.3, 17.5\pm0.3,
18.0\pm0.4$ within the same solid angles in logarithmic units of ${\rm
  GeV \, cm^{-2}}$ (see Table~\ref{tuc4_table}).  Upper limits on the
J-factor of Tuc~V are listed in Table~\ref{tuc5_table}.  We also set
upper limits of $\log_{10}{D}<18.0, 18.6$ at solid angles of
$0.2\degree,0.5\degree$ for Tuc~V.  Because of the small sample size
for Tuc~V the upper limits are not very meaningful, whereas Gru~II has
one of the smallest J-factors given the computed upper limit.  The
J-factor for Tuc~IV is not very large compared to other ultra-faint
dwarfs mostly due to its large size, but agrees with scaling relations
based on its velocity dispersion, distance, and half-light radius
\citep{ps19}.

Using the sizes, distances, and luminosities of dwarfs with measured
kinematics, \citet{ps19} made predictions for dwarfs lacking stellar
kinematics.  They predicted $\log{J}=18.4, 18.1,18.9$ for Gru~II,
Tuc~IV, and Tuc~V, respectively.  The prediction for Tuc~IV agrees
with our J-factor measurement whereas the prediction for Gru~II is
much larger than our upper limit, emphasizing the importance of
stellar kinematics for determining accurate J-factors.

Tuc~IV is one of the four dwarfs with an excess ($\sim2\sigma$) of
gamma-rays detected by Fermi \citep{albert17}.  Two of the other
dwarfs with an excess, Ret~II and Tuc~III, have large and small
J-factors, respectively, relative to the ultra-faint population as a
whole \citep{ps19}.  The Tuc~IV J-factor falls between these two.
\citet{albert17} noted that there was no correlation between the
measured (or predicted) J-factor and the flux upper limit for the
dwarf galaxy population, and we find that Tuc~IV continues this trend.
With a measured J-factor, Tuc~IV will be a useful addition to any
future searches of dark matter annihilation or decay.

\subsection{Assessing the Possibility of Tidal Stripping}

Since their discovery 15 years ago, extensive speculation has centered
on the effects of tides on the ultra-faint dwarfs
\citep[e.g.,][]{zucker06,coleman07,penarrubia08,munoz10,br11a,deason12,lokas12,collins17,mp19}.
Tidal stripping is particularly important for interpreting the masses
of these systems; a galaxy that has recently experienced a tidal shock
may have a velocity dispersion that is not representative of its
current mass \citep[e.g.,][]{kuepper17}, and over many orbits
stripping can potentially reduce the mass of a galaxy by an order of
magnitude or more \citep[e.g.,][]{kravtsov04,penarrubia08}.  Tidal
stripping may also have significant implications for the number of
satellite galaxies and their radial distribution relative to the Milky
Way \citep[e.g.,][]{gk17} as well as for the luminosity-metallicity
relationship in the dwarf galaxy regime \citep[e.g.,][]{kirby13b}.


For individual systems, the minimum observed velocity dispersion for
dwarf galaxies has been decreasing with new discoveries and improved
kinematics \citep[e.g.,][]{collins17,koposov18}, and the previous
examples of systems with unusually cold internal kinematics for their
luminosities \citep{kirby13,simon17,caldwell17} all have substantial
evidence for tidal stripping.  In particular, Segue~2 ($\sigma <
2.2$~\kms; \citealt{kirby13,simon19}) is at least 0.6~dex more
metal-rich than would be expected from its luminosity, Tucana~III
($\sigma < 1.2$~\kms; \citealt{simon17}) exhibits tidal tails
\citep{drlica15,li18b}, and the large radius of Crater~II ($\sigma =
2.7 \pm 0.3$~\kms; \citealt{caldwell17}) implies significant tidal
mass loss on its derived orbit \citep{sanders18,fu19}.  Gru~II does
not obviously fit this trend, with an upper limit to its velocity
dispersion comparable to that for Tuc~III and Segue~2 but no detected
tidal tails\footnote{Note that Tuc~III is on an extremely radial
  orbit, which will result in more prominent tails than a comparable
  satellite on a less eccentric orbit.}, as well as a metallicity that
is compatible with the dwarf galaxy metallicity-luminosity
relationship at its present luminosity.  However, using the $1.6
\times 10^{12}$~\msun\ Milky Way mass model from \citet{carlin18} and
assuming that (1) the tidal radius is equal to the Jacobi radius as
defined by \citet{bt08} and (2) no mass is present beyond the
half-light radius, the tidal radius of Gru~II at its current location
is $296~(\sigma/2.0~\kms)^{2/3}$~pc.  Thus, even if the true velocity
dispersion is equal to the $2\sigma$ upper limit determined in
Section~\ref{sec:kin}, $r_{\rm tidal}/r_{1/2} = 3.1$, implying that up
to $\sim10$\% of the stars in Gru~II could be vulnerable to
stripping.\footnote{Our assumption here that the dynamical mass is
  equal to the mass within the half-light radius is of course
  extremely conservative.  In principle the total mass of the system
  could be much larger if Gru~II is still embedded in an extended dark
  matter halo, but we have no way of measuring that mass.  The
  calculations here demonstrate that tidal stripping is possible, but
  are not sufficient to establish that it is definitely occurring.}
If the actual velocity dispersion is significantly smaller, even more
stars could be stripped, especially as Gru~II approaches the
pericenter of its orbit.

Under the same conservative assumptions as for Gru~II above, we
calculate a tidal radius for Tuc~IV of 481~pc ($3.8 r_{1/2}$).  Thus,
even if Tuc~IV lacks an extended dark matter halo very few of its
stars are beyond its tidal radius and likely to be tidally stripped by
the Milky Way.  As discussed in \S~\ref{sec:lmc_collision}, whether
the central portion of Tuc~IV remained tightly bound during its recent
collision with the LMC depends on if the system retained a massive
halo until that time.  Even if so, it is certainly possible that the
outer regions of the dwarf, as well as much of its dark matter, could
have been removed by that interaction.

For Tuc~V we find a tidal radius of $386~(\sigma/7.4~\kms)^{2/3}$~pc.
The compact size of Tuc~V makes it more resilient to tidal stripping.
Unless its velocity dispersion is less than 1.0~\kms, its tidal radius
is at least $3 r_{1/2}$.  Our photometric analysis of Tuc~V determined
an elongated shape with an ellipticity of $e = 0.51^{+0.09}_{-0.18}$.
Some authors have suggested that such shapes are a signature of tidal
stripping \citep[e.g.,][]{coleman07,martin08,deason12}, but more
detailed calculations indicate that tidal disruption generally does
not induce significant ellipticities \citep{munoz08}.  Since its orbit
has likely not brought it within $\sim30$~kpc of the Milky Way, we
conclude that Tuc~V has probably not suffered any tidal stripping.

\subsection{Magellanic Association}
\label{mc_assoc}

A number of authors have recently examined the question of which
nearby dwarf galaxies might actually be satellites of the Magellanic
Clouds \citep{jethwa16,sales17,simon18,kallivayalil18,pardy19,eb19}.
In particular, \citet{jethwa16}, \citet{sales17}, and
\citet{kallivayalil18} provided predictions for the kinematics of our
targets if they are LMC or SMC satellites.  While the radial
velocities we measure fall within the 68\% confidence intervals
computed by \citet{jethwa16} for all three objects, as already noted
by \citet{pl19} the proper motions do not agree with the
\citeauthor{jethwa16} predictions.  We find that the radial velocity
and proper motion in the Galactic longitude direction of Tuc~V both
strongly disagree with the predictions of \citet{sales17}, reinforcing
our conclusion from \S~\ref{sec:lmc_collision} that it is not a
Magellanic satellite.  For Tuc~IV, the observed radial velocity and
proper motion in Galactic latitude are in reasonable agreement with
\citet{sales17}, but the proper motion in Galactic longitude differs
by $\sim200$~\kms.  Given the recent interaction between Tuc~IV and
the LMC, our calculations in \S~\ref{sec:lmc_collision} represent a
more complete assessment of its membership in the Magellanic group.
\citet{sales17} do not provide predictions for Gru~II because they
judge it to have too low a probability of association with the
Magellanic Clouds.  Relative to the predictions of
\citet{kallivayalil18}, our measurements for Gru~II are the closest to
matching, but none of our targets have motions that agree with the
predicted values in all three dimensions.


\subsection{A Tucana Group?}

As mentioned in Section~\ref{intro}, the satellites Tuc~II, Tuc~IV,
and Tuc~V are quite close together, and could potentially be the
remnant of a dwarf galaxy group \citep{drlica15}.  However, the radial
velocities and proper motions of Tuc~IV and Tuc~V place them on quite
different orbits from each other, as well as from Tuc~II.  There is
therefore no evident connection between the three satellites, and
their current proximity is largely coincidental, as is the case for
the close pair of Car~II and Car~III.  It is possible, though, that
the gravitational influence of the LMC has played a role in bringing
these systems closer together.

\subsection{Connection to Other Halo Substructures}

\citet{koposov19} used Gaia DR2 data to trace the Orphan Stream
\citep{belokurov07b} far into the southern hemisphere, finding that it
passes very close to Gru~II.  At the point of closest approach, the
angular separation between the two is $\sim1$\degr, and the proper
motion of the stream stars is also very similar to that of Gru~II.
However, the accompanying modeling by \citet{erkal19} determines a
heliocentric velocity for the stream of $\sim-200$~\kms\ at this
position, which is offset by $\sim90$~\kms\ from the velocity we
measure for Gru~II.  Moreover, the stream stars are $\sim10$~kpc
closer than Gru~II \citep{koposov19}.  The difference in distance and
velocity between Gru~II and the Orphan Stream likely rules out any
connection between the two \citep{martinez19}.  In our spectroscopic
data set we find two stars close to the expected velocity and proper
motion of the stream, DES~J220441.57$-$462244.8 and
DES~J220458.78$-$462710.6.  Both stars have colors and magnitudes
consistent with being RGB stars at a distance modulus $\sim0.4$~mag
smaller than that of Gru~II, and CaT metallicities of $\feh \approx
-2$ assuming that distance.  These stars can be added to the sample of
spectroscopic members of the Orphan Stream being assembled by the
S$^{5}$ collaboration \citep{li19}.

\section{SUMMARY AND CONCLUSIONS}
\label{conclusions}

We have presented the first spectroscopic analysis of the Milky Way
satellites Gru~II, Tuc~IV, and Tuc~V.  Using medium-resolution
Magellan/IMACS spectroscopy, we identified 21 member stars in Gru~II,
11 in Tuc~IV, and 3 in Tuc~V.  We used these data to measure the
radial velocity, mean metallicity, and proper motion of each system.
We determined the velocity dispersion of Tuc~IV, but were only able to
derive upper limits on the velocity dispersions of Gru~II and Tuc~V.
None of the three objects have a detectable metallicity spread in the
existing data.

Based on its low metallicity and large size, we conclude that Gru~II
is most likely a dwarf galaxy.  The same characteristics for Tuc~IV,
along with the large dynamical mass indicated by its velocity
dispersion, identify it as a dwarf galaxy as well.  Because of the
small number of bright member stars in Tuc~V our constraints on its
velocity dispersion, mean metallicity, and metallicity dispersion are
very weak.  Combined with its small size, we are unable to draw any
significant conclusions about whether Tuc~V is a dwarf galaxy or
globular cluster.

We employed the three-dimensional velocities of the three satellites
to compute their orbits around the Milky Way.  Gru~II and Tuc~IV are
on eccentric orbits with pericenters of $\sim20-30$~kpc and apocenters
of $\sim50-60$~kpc, similar to the orbits of other nearby ultra-faint
dwarfs \citep[e.g.,][]{simon18,fritz18}.  In contrast, the orbit of
Tuc~V likely extends beyond a distance of 100~kpc, and there is a
non-negligible chance that Tuc~V is on its first infall to the Milky
Way.  All three systems are currently approaching pericenter.  By
projecting the orbits backward in time, we discovered that Tuc~IV
recently collided with the LMC, with an impact parameter of
$\sim4$~kpc.  Based on their orbits and internal kinematics, we
conclude that Gru~II could have suffered modest tidal stripping by the
Milky Way, Tuc~IV could have been stripped by the LMC, and Tuc~V is
unlikely to have been stripped.

\acknowledgements{This publication is based upon work supported by the
  National Science Foundation under grant AST-1714873.  JDS was also
  partially supported by Program number HST-GO-14734, provided by NASA
  through a grant from the Space Telescope Science Institute, which is
  operated by the Association of Universities for Research in
  Astronomy, Incorporated, under NASA contract NAS5-26555.  Some of
  this work was carried out during a stay at the Kavli Institute for
  Theoretical Physics, which is supported in part by the National
  Science Foundation under Grant No. NSF PHY-1748958, for the program
  \emph{The Small-Scale Structure of Cold(?) Dark Matter}.  TSL was
  supported by NASA through Hubble Fellowship grant HST-HF2-51439.001
  awarded by the Space Telescope Science Institute, which is operated
  by the Association of Universities for Research in Astronomy, Inc.,
  for NASA, under contract NAS5-26555.  We thank Dan Kelson for many
  helpful conversations regarding IMACS data reduction and Petchara
  Pattarakijwanich for early contributions to the IMACS reduction
  procedures, as well as the anonymous referee for comments that
  helped improve the paper.  This research has made use of NASA's
  Astrophysics Data System Bibliographic Services. Contour plots were
  generated using \code{corner.py} \citep{corner}.

  Funding for the DES Projects has been provided by the
  U.S. Department of Energy, the U.S. National Science Foundation, the
  Ministry of Science and Education of Spain, the Science and
  Technology Facilities Council of the United Kingdom, the Higher
  Education Funding Council for England, the National Center for
  Supercomputing Applications at the University of Illinois at
  Urbana-Champaign, the Kavli Institute of Cosmological Physics at the
  University of Chicago, Financiadora de Estudos e Projetos, Funda{\c
    c}{\~a}o Carlos Chagas Filho de Amparo {\`a} Pesquisa do Estado do
  Rio de Janeiro, Conselho Nacional de Desenvolvimento Cient{\'i}fico
  e Tecnol{\'o}gico and the Minist{\'e}rio da Ci{\^e}ncia e
  Tecnologia, the Deutsche Forschungsgemeinschaft and the
  Collaborating Institutions in the Dark Energy Survey.  The DES
  participants from Spanish institutions are partially supported by
  MINECO under grants AYA2012-39559, ESP2013-48274, FPA2013-47986, and
  Centro de Excelencia Severo Ochoa SEV-2012-0234, some of which
  include ERDF funds from the European Union. This material is based
  upon work supported by the National Science Foundation under Grant
  Number (1138766).

  The Collaborating Institutions are Argonne National Laboratory, the
  University of California at Santa Cruz, the University of Cambridge,
  Centro de Investigaciones Energ{\'e}ticas, Medioambientales y
  Tecnol{\'o}gicas-Madrid, the University of Chicago, University College
  London, the DES-Brazil Consortium, the Eidgen{\"o}ssische Technische
  Hochschule (ETH) Z{\"u}rich, Fermi National Accelerator Laboratory,
  the University of Edinburgh, the University of Illinois at
  Urbana-Champaign, the Institut de Ci{\`e}ncies de l'Espai (IEEC/CSIC),
  the Institut de F{\'i}sica d'Altes Energies, Lawrence Berkeley National
  Laboratory, the Ludwig-Maximilians Universit{\"a}t and the
  associated Excellence Cluster Universe, the University of Michigan,
  the National Optical Astronomy Observatory, the University of
  Nottingham, The Ohio State University, the University of
  Pennsylvania, the University of Portsmouth, SLAC National
  Accelerator Laboratory, Stanford University, the University of
  Sussex, and Texas A\&M University.
}

{\it Facilities:} \facility{Magellan:I (IMACS)}


\bibliographystyle{apj}

\clearpage

\end{document}